Ben-Gurion University of the Negev
Faculty of Engineering Science
Department of Information Systems Engineering

# The Security of Organizations and Individuals in Online Social Networks

THESIS SUBMITTED IN PARTIAL FULFILLMENT OF THE REQUIREMENTS
FOR THE M.Sc. DEGREE

**Submitted by**
Aviad Elyashar
Department of Information Systems Engineering
Ben-Gurion University of the Negev
Tel: +972-52-6056439
E-mail: aviad.elishar@gmail.com

**September 21, 2015**

Ben-Gurion University of the Negev
Faculty of Engineering Science
Department of Information Systems Engineering

# The Security of Organizations and Individuals in Online Social Networks

THESIS SUBMITTED IN PARTIAL FULFILLMENT OF THE REQUIREMENTS
FOR THE M.Sc. DEGREE

By: Aviad Elyashar (aviad.elishar@gmail.com)

Supervised by: Prof. Yuval Elovici (elovici@post.bgu.ac.il)

Dr. Michael Fire (fire@cs.washington.edu)

**September 21, 2015**

# Abstract

Online social networks (OSNs) have gained great popularity in recent years, having become an integral part of our daily lives and an indispensable dimension of the Internet. Users worldwide spend a significant amount of their time on OSNs, which have enabled them to create new connections with others based on shared interests, activities, and ideas, as well as maintain connections from the past.

Along with the positive aspects of OSNs, users are faced with some uniquely troublesome issues; chief among these issues are privacy and security. Security breaches often result when users write, share, and publish personal information about themselves, their friends, and their workplaces online, not realizing the information they post is available to the public and can be used to aid malicious hackers. Commonly shared personal information may include, but is not limited to personal photographs, date of birth, religious affiliations, personal interests, and political views.

In our study, we attempted to emphasize the acute problems inherent on OSNs that leave both employees and organizations vulnerable to cyber-attacks. In many cases, these attacks include the use of socialbots, which among other things, can spread spam and malware, and "phish" for login and password information. These malicious attacks may result in identity theft, fraud, and loss of intellectual assets and confidential business information.

The serious privacy and security problems related to OSNs are what fueled two complementary studies as part of this thesis. In the first study, we developed a general algorithm for the mining of data of targeted organizations by using Facebook (currently the most popular OSN) and socialbots. By friending employees in a targeted organization, our active socialbots were able to find new employees and informal organizational links that we could not find by crawling with passive socialbots. We evaluated our method on the Facebook OSN and were able to reconstruct the social networks of employees in three distinct, actual organizations. Furthermore, in the crawling process with our active socialbots we discovered up to 13.55% more employees and 22.27% more informal organizational links in contrast to the crawling process that was performed by passive socialbots with no company associations as "friends".

In our second study, we developed a general algorithm for reaching specific OSN users who declared themselves to be employees of targeted organizations, using the topologies of organizational social networks and utilizing socialbots. We evaluated the proposed method on targeted users from three actual organizations on Facebook, and two actual organizations on the Xing OSN (another popular OSN platform). Eventually, our socialbots were able to reach specific users with a success rate of up to 70% on Facebook, and up to 60% on Xing.

The results from both studies demonstrate the dangers associated with OSNs. We believe that raising awareness regarding privacy issues among all entities of OSNs — users, organizations, and OSN operators — as well as developing preventive tools and policies, may help to resolve this widespread, critical situation, and better protect OSN users' privacy and security.

## Keywords

## Acknowledgment

I would like to express my sincere gratitude to my supervisors, Prof. Yuval Elovici and Dr. Michael Fire, for their continuous support of my M.Sc. studies and research —and especially for their patience, motivation, enthusiasm, and immense knowledge. Their guidance has helped me throughout my research and in the construction of this thesis.

In addition, I would like to thank Ms. Marisa Timko for proofreading this thesis. Her editing expertise and continuous helpful advice has helped me bring this thesis to completion.

# Table of Contents

# Table of Figures

# Table of Tables

# 1. Introduction

For more than a decade, the Internet has played an increasingly major role in our lives. The Internet has amassed billions of users all over the world [1]. Users access the Internet by means of various digital devices such as computers, tablets, and smartphones for personal and business needs [2]. Moreover, the diversity of the activities users can carry out on the Internet platform is quite extensive. For example, they are able to surf the web and play games [3], purchase goods [4], schedule appointments, download files and software, pay bills, conduct meetings, and even read medical information [5].

One of the most successful sectors on the Internet that has gained great popularity in recent years is the platform of online social networks (OSNs) [6], [7]. Popular OSNs like Facebook,[1] LinkedIn,[2] MySpace,[3] and Xing[4] allow Internet users to create user accounts and maintain connections with others.

The reasons for the growing popularity of OSNs are varied. One of the reasons is based on the fact that OSNs enable users to create new connections with others based on shared interests, activities, and ideas, etc., as well as providing a forum for maintaining connections from the past [6]. Other reasons are based on the fact that OSNs are multi-optional platforms for self-expression [8]. They provide a forum through which users can publish statuses and photos, play games, initiate romantic relationships, chat, and so on [9].

Along with the numerous benefits that OSNs provide —such, as maintaining relationships, finding new colleagues, and promoting businesses —there are also many threats that may jeopardize the security of OSN users, as well as that of their workplaces [10], [11].

For example, today, many OSN users display the name of their workplace on their profile account [11], [12], [13], [14]. Malicious users may take advantage of this by recording such information for later use [15]. After gathering enough information regarding the employees of the targeted organization, they can use it to construct the organization's structure [11]. After the reconnaissance process ends, they may utilize the sensitive information to perform malicious attacks on key-role employees. These attacks could result in fraud, as well as the loss of intellectual assets, and confidential business information for targeted organizations [14].

In recent years, socialbot attacks were found to be preferred by several adversaries [1], [8], [12], [13], [14], [16]. Socialbots, which are also known as sybils [17], are defined as automatic or semi-automatic computer programs that control OSN

---

[1] www.facebook.com
[2] www.linkedin.com
[3] www.myspace.com
[4] www.xing.com

accounts and perform human behaviors, such as sending friend requests and messages, etc. [1], [18].

This thesis is a combination of two complementary studies which are focused upon socialbot attacks. In the first study, we empirically measure the additional amount of information we could find on a targeted organization by using socialbots. In order to assess the amount of additional information, we developed a general algorithm for the mining of data of targeted organizations by using the Facebook OSN and socialbots (see Section 3.2.1). First, we crawled with a Facebook passive socialbot with no friends. The public information we gathered from this user was defined as the minimal public information that a user can gain within an OSN. Afterward, we crawled once again, but this time with a socialbot account that we created within the Facebook OSN. This socialbot befriended several employees in a targeted organization prior to the crawling process. By gaining the information regarding the connections and the role of these employees from the targeted organization, our socialbot was able to construct the organizational structure, as well as to find hidden connections and employees that we were not able to find by crawling with the passive socialbot. We evaluated our method on the Facebook OSN and were able to construct the social networks of three separate organizations. Moreover, we discovered up to 13.55% more employees and up to 22.27% more informal organizational connections during the crawling process using the active socialbots in contrast to the passive ones without any friends (see Section 4.1). An informal connection is defined as a social connection of individuals without formal structure. This connection is based often on friendship, ethnicity, neighborhood, etc. [19]

In the second study, we continued our research by altering the study's focus from a targeted organization to a targeted organization's employees. We developed an algorithm for reaching specific OSN users who declared themselves to be employees of targeted organizations, by using topologies of organizational social networks, and socialbots (see Section 3.2.2). We evaluated the proposed method on targeted users from three actual organizations on Facebook, and two actual organizations on the Xing OSN. Eventually, our active socialbots were able to reach specific users with a success rate of 50%, 70% and 40% on Facebook, and 20%, and 60% on Xing (see Section 4.2).

Lastly, we present suggestions and recommendations to prevent and detect socialbots from stealing great assets from targeted organizations and employees (see Section 7).

Our results from both studies demonstrate that dangers within OSNs are prevalent. A user's personal information may be disclosed to malicious third parties. Adversaries can use this information for reconnaissance and then to attack users on several platforms, such as within OSNs, in emails, etc. As a result, we believe that when using OSNs, users, organizations, and operators should be more careful and aware of the privacy issues. Furthermore, developing preventative tools, as well as restricting policies, may help to improve OSN users' privacy and security.

## 1.1.     Contributions

This thesis was carried out in order to emphasize the serious threats that may be posed for organizations and employees in this time of the growing popularity of OSNs. Organizations and employees may find themselves in great danger on OSN platforms.

Specifically, this thesis offers the following contributions: First, to the best of our knowledge, we are the first to differentiate between public organizational information crawled by a passive socialbot with no friends, and crawling carried out by an active socialbot which gained several employees as friends in a targeted organization (see Section 4.1). Attaining such information awarded us the opportunity to extract a network of informal social relationships and to find new friendships and employees that we could not find before.

Second, in contrast to several studies that discuss attempts to reach users through OSNs without making any distinction between them [1], [8], we chose to focus on employees in an organization to define as targets whom adversaries would be interested in infiltrating.

Third, as opposed to similar studies that focused on reaching OSN users, this thesis also enhances the focus on organizations. Our studies were not defined as solely user-oriented, but also as organization-oriented. We introduced the privacy issues within OSNs that threaten to endanger employees as well as organizations.

Fourth, this study is the first to evaluate socialbot attacks on organizations in more than one OSN: Facebook, and Xing.

Lastly, to the best of our knowledge, this is the first study which offers a generic algorithm for reaching specific users within OSNs by means of socialbots. No other studies attempted to do this before.

## 1.2.    Publications

This thesis was a product of a fruitful collaboration between the author of the dissertation along with Dr. Michael Fire, Mr. Dima Kagan, and Prof. Yuval Elovici.

It is worth mentioning that the work presented herein consists of research studies that have been published in international conferences, workshops, and journals. In particular, the characterization study presented in Section 3 and 4, led to the following publications:

- **Aviad Elyashar**, Michael Fire, Dima Kagan, and Yuval Elovici. "Guided socialbots: Infiltrating the social networks of specific organizations' employees", AI Communications, IOS Press, 2014.

- **Aviad Elyashar**, Michael Fire, Dima Kagan, and Yuval Elovici. "Homing Socialbots: Intrusion on a specific organizations employee using Socialbots", SNAA - Social Network Analysis in Applications 2013, Niagara Falls, Ontario, Canada, 2013.

- **Aviad Elyashar**, Michael Fire, Dima Kagan, and Yuval Elovici. "Organizational Intrusion: Organization Mining using Socialbots", 2012 International Conference on Social Informatics (SocialInformatics), Lausanne, Switzerland.

Other publications which focused on developing tools for improving privacy among users resulted in the following:

- Michael Fire, Dima Kagan, **Aviad Elyashar**, and Yuval Elovici, "Friends or Foe? Fake Profile Identification in Online Social Networks," Springer Journal of Social Network Analysis and Mining (SNAM), 2014, In Press.

- Dima Kagan, Michael Fire, **Aviad Elyashar**, and Yuval Elovici, "Facebook Applications Installation and Removal- Temporal Analysis", The Third International Conference on Social Eco-Informatics (SOTICS), Lisbon, Portugal, November 2013.

- Michael Fire, Dima Kagan, **Aviad Elyashar** ,and Yuval Elovici " Social Privacy Protector - Protecting Users' Privacy in Social Networks" SOTICS 2012 : The Second International Conference on Social Eco-Informatics

## 1.3.    Organization

The remainder of this thesis is organized as follows: In Section 2 we provide an extensive overview of literature that focuses on similar issues to those which we discuss in the current study. Section 3 describes the experimental framework and methods we used in order to carry out our two experiments. Furthermore, this section includes the algorithms we developed for the mining of data of organizations as well as reaching specific employees, the obstacles we faced during these experiments, and the datasets we used in order to evaluate our methods. Section 4 presents our numeric results. Section 5 includes ethical considerations that arose during this study. Section 6 presents a discussion regarding the results, and Section 7 presents our conclusions and future research directions.

## 2. Literature Overview

In this section, we describe several studies related to the issues we focus on in the thesis to provide helpful background information and additional insights.

First, we concentrate on OSNs. We attempt to understand what the term "*online social network*" means, and explore the reasons for OSN success. Then, we show how much diversity OSNs provide for their users. Later, we explore the threats involved when using OSNs and the privacy problems that exist within them.

Moreover, we outline studies involving trust issues, the crawling of OSNs, clustering methods for analyzing organizations, the use of socialbots—including definitions, studies that used socialbots for attacking, and the identification of socialbots.

Finally, we review security tools that protect OSN users.

### 2.1. Online Social Networks

In this section, we discuss studies that focused on OSNs.

#### 2.1.1. Online Social Networks Definition

In 2007, Boyd et al. [6] defined OSNs. They defined them as web-based services that provide individuals with an infrastructure to create a public or private profile account within a bounded system. This system enables a user to establish a profile that includes a list of other users with whom he or she shares connections. Typically, these other users are defined as the user's friends. Moreover, users can view and traverse their list of connections and those made by others within this system.

#### 2.1.2. Reasons for Attraction

The success of OSN usage among users may be attributed to several factors. Users want to stay connected with their surroundings, and by registering with OSNs, they can stay updated about their friends' whereabouts and maintain closer relationships with them [7].

Others utilize OSNs to promote businesses. For years, self-employed individuals such as barbers, plumbers, or accountants have been using OSNs to publicize their businesses in order to increase their customer base.

Today, users even rely upon OSNs in order to initiate romantic relationships [9].

Malicious users are also attracted to OSNs as these platforms are deemed fertile ground for running astroturf campaigns. The goal in conducting such campaigns is to spread misinformation and propaganda in order to bias public opinion [1], [20]. Other attackers use OSNs for spreading spam [8] and even to influence users with manipulation [18].

### 2.1.3.  Diversity

There is a great diversity among OSNs; they offer a variety of networks to correlate with different aspects of users' lives. The largest OSN in the world is Facebook with more than one billion users [21], [22]. However, there are many more OSNs which help users to connect based on shared interests, political views, or common activities. These include LinkedIn, one of the world's largest professional networks; Xing, a European social business network for business professionals; Academia.edu,[5] a social networking site for academics and researchers; Athlinks,[6] a social networking website aimed at competitive endurance athletes; and many more.

### 2.1.4.  The Privacy Problem in Online Social Networks

One of the major problems in OSNs revolves around the privacy issue. Most of the OSNs allow Internet users to create a user profile in order to present themselves in the social networks, and to initiate connections with others. This procedure enables positive actions, such as finding new friends and colleagues based on common interests and activities, political views, sexual attraction, etc.; however, there are also many negative implications pertaining to this procedure. Most of the time in order to present themselves on OSNs, users upload pictures and publish private and personal information about themselves, such as their name, age, gender, sexual orientation, preferred establishments, phone numbers, address,  etc.

For example, in 2005, Gross et al. [23] discussed patterns of information exposure in OSNs and their privacy implications. They examined the online behavior of more than four thousand university students in a popular social network in terms of the amount of disclosed information by users and their implantation of privacy settings. Eventually, they concluded that only small numbers of students changed their privacy preferences from the default privacy preferences, which expose too much information regarding users.

In 2006, Barnes [24] discussed privacy issues within OSNs. She found OSN sites to be like a magnet, which attracts many American youngsters. She indicated that teenagers using OSNs freely shared personal information. This personal information attracts sexual predators. In general, she declared that many people may not be aware of the fact that their privacy is in danger and that they are not doing anything to protect their personal information. Such sensitive information can come in the form of home address, phone numbers, pictures, etc. The solution Barnes arrived at is not a simple one; in order to tackle issues which can result in the loss of teens' Internet privacy, a keen awareness and effort on all levels of the society must be brought about and executed.

In 2009, Lindamood et al. [25] argued that some of the information revealed on social networks is private and it is possible that corporations could use learning algorithms on the released data in order to predict undisclosed private information. They found that removing trait details and friendship links is the best way to reduce classifier accuracy.

---

[5] http://academia.edu
[6] http://athlinks.com

### 2.1.5. OSN Threats

Along with the numerous benefits that OSNs provide, such as maintaining relationships, finding new colleagues, and promoting businesses, there are also many threats that may jeopardize the security of OSN users, as well as that of their workplaces. These threats can be divided into three major classifications.

### 2.1.5.1. Individual Security

The first is that threats pose a danger to individual security. Today, many OSN users are unaware of the serious privacy issues that accompany the use of OSNs [10]. Users often share personal data on OSNs without realizing the short-term or long-term consequences of such information-flow [24], [26]. Gathered data that discloses personal and sensitive information about users may cause security risks, including: identity theft [8]; inference attacks [27]; spreading spam [10], [28]; privacy threats [1], [29]; malware [30]; fake profiles or sybils [31], [32]; socialbots [1], [33]; and sexual harassment [34], [35].

### 2.1.5.2. Business Security

The second threat is that to business security. Malicious users may engage in industrial espionage by creating fake profiles or bots in order to connect to users who are key employees in targeted organizations. By so doing, the hackers gain access to monitor the information that users disclose [11], [36]. An exposed user's information pertaining to an organization may result in a loss of intellectual assets and confidential business information, as well as sensitive business data, which could make the company vulnerable to stock market manipulation and cybercrimes – ultimately setting the organization back hundreds of millions of dollars every year [37]. Moreover, malicious users may spread rumors regarding the targeted organization that could result in serious reputational damage, without the ability to track the source of the rumors [38], [39].

### 2.1.5.3. National Security

The third major classifications of threats are those posed to national security. Soldiers may inadvertently disclose confidential operational information to their friends through OSNs [40]. The enemy may collect these national secrets, like undisclosed locations, and use this information as an advantage in the future. Moreover, hackers may use virtual identities in order to spread propaganda by connecting with key users in the OSN to demoralize the opponent's society. Furthermore, the enemy may use the exposed information to run astroturf campaigns for spreading propaganda or misinformation regarding important issues such as US political elections [1], [41], [42]. Moreover, in recent years, some terrorists around the world utilized OSNs as a covert communication channel by which they operated to better organize and coordinate dispersed activities [43].

### 2.1.6. Targeted Online Social Networks

During our study, we utilized the Facebook and Xing OSNs to conduct our experiments. In this section, we provide a brief overview of each one of these networks.

#### 2.1.6.1. Facebook

With more than 1.49 billion monthly active users as of June 30, 2015, Facebook stands out among all other popular OSNs in the world. According to Facebook, there are 1.31 billion active monthly users of its mobile products. On average, Facebook has 968 million active daily users, and 83.1% of them are outside the U.S. and Canada [21]. The average Facebook user has around 190 friends [44].

According to Facebook's estimations [45], 8.7% of its accounts are defined as fake, meaning that Facebook includes tens of millions of fake accounts. Moreover, 4.8% of Facebook accounts are defined as duplicate accounts—ones that users maintain in addition to their principal accounts. Furthermore, 2.4% of Facebook users are defined as user-misclassified accounts, referring to users who have created personal profiles for a business, organization, or non-human entity, such as a pet. Of Facebook's fake accounts, 1.5% are defined as undesirable accounts, i.e., belonging to malicious users. Undesirable accounts are defined as fake accounts which were created with the intent of being used for purposes that violate Facebook's terms of service, such as spamming or distributing other malicious links and content.

Regarding security issues, Facebook uses the FIS (Facebook Immune System) [46] and provides an additional privacy settings tool which enables users to edit their profile and decide which information will be accessible to others. However, this tool matched Facebook users' expectations only 63% of the time. Furthermore, the Facebook privacy settings tend to expose content to more users than expected [47].

#### 2.1.6.2. Xing

Xing is a social network for business professionals. It was founded in Hamburg, Germany in 2003 and has been publicly listed since 2006. Xing has around 8.8 million members worldwide, 8.4 million of whom are based in German-speaking countries [48]. Most Xing users use this OSN to promote their businesses, boost their career, or find a job.

Furthermore, Xing provides a suitable platform for professionals to meet, find jobs, connect with colleagues, collaborate, share new ideas, etc. [48] Regarding security issues, Xing provides information about how active a given user is (based on the frequency of a user's visits)—information which is referred to as *activity*. With this data, we can assess the activity level of Xing users.

Moreover, Xing keeps track of the number of unconfirmed users, i.e., users who did not confirm a specific user. When a user is unconfirmed one hundred times, Xing prevents the user from sending additional friend requests [49]. To the best of our knowledge Xing does not publish statistics regarding the number of fake users within the network.

## 2.2.    Trust

OSNs in general as well as socialbots, base their success on trust. Users tend to trust other users based on several properties such as gender, occupation, connections, etc. As a result, users increase their friend list and their usage of OSNs.

In 2007, Dwyer et al. [50] focused on the issues of trust and privacy within OSNs. The researchers compared two well-known OSNs, Facebook and MySpace, and came to the conclusion that with regard to online interaction, trust is less important in the construction of new relationships than it is in face-to-face encounters. Furthermore, they were able to show that trust along with the willingness to share information do not automatically translate into new social interactions online. They found that Facebook members revealed more information than MySpace members, however, the latter were found to be more likely to extend online relationships beyond the bounds of the social networking site, despite weaker trust results. Malicious users take advantage of this fact in order to spread themselves in these sites and find victims.

In 2010, Ryan et al. [51] conducted the Robin Sage Experiment. In their experiment they created a false identity named Robin Sage and activated the identity on various social networking websites. Robin joined networks in order to influence users. In this case, the influence was reflected in the ability to gain the trust of other users, while having them share sensitive information with the false identity. The main points of interest observed by Ryan et al. were the ability to exploit strangers' level of trust based on attributes such as occupation, gender, education, and friends. This experiment was conducted over the span of a month and ultimately, Robin gained hundreds of friends on various social networking sites. Among her friends were executives at government entities such as the NSA (National Security Agency), military intelligence groups, Fortune Global 500 corporations, and so on. Moreover, the false identity even received gifts, government and corporate jobs offers and even invitations to speak at several security conferences. Ryan et al. attributed the success of this experiment to several factors. First, the security industry is dominated by males which makes the female presence a rarity in this sector. For this reason, they deliberately chose a young, attractive female for the fabricated identity. Secondly, Robin's education and credentials combined with her certifications deemed her an experienced security professional–driving others to connect with her for career opportunities, knowledge expansion, and so on.

## 2.3.    Crawling Social Networks

In recent years there have been several studies conducted which focused on crawling OSNs for many reasons. The crawling process is often done for the retrieval of large amounts of information from OSNs.

For example, in 2007, Chau et al. [52] emphasized how easy it is to retrieve information from OSNs. Chau et al. described their implementation of crawlers for OSNs. These crawlers were able to visit a total of approximately 11 million auction users.

In 2010, Kwak et al. [53] crawled a considerable amount of Twitter[7] information. They gathered 41.7 million user profiles, 1.47 billion social relations, 4,262 trending topics, and 106 million tweets.

In 2012, Fire et al. [11] analyzed different organizations through data mining techniques. The analysis process was based on information from organizations' employees revealed on Facebook, LinkedIn, Google search results, the company's web page and other publicly available sources. To accomplish their goal, they designed and built a novel web crawler. This web crawler extracted information about the given network such as informal social relationships of employees of a given target organization. In contrast with standard crawlers, which were found to be insufficient in performing data collection as they collected many irrelevant profiles and skipped Facebook users who worked in a target organization, the newly designed web crawler optimized data collection from users associated with a specific group or organization. Fire et al. collected publicly available data from six well-known hi-tech companies on three different scales using the new web crawler.

In 2013, Stern et al. [54] introduced the Target Oriented Network Intelligence Collection (TONIC) problem. The goal of this problem was to find OSN profiles that contain information regarding a given target using automated crawling. Stern et al. attempted to retrieve information regarding a specific target while avoiding further crawling. They defined TONIC as a heuristic search problem and solved it with Artificial Intelligence techniques. They evaluated their problem on data set from Google+ OSN that included 211,000 profiles with 1.5 million links between them. Eventually, Stern et al. found that the Bayesian Promising heuristic significantly outperforms other heuristics.

## 2.4.    Clustering Methods for Analyzing Organizations

In 1979, Tichy et al. [55] described a technique for analyzing organizations using a network that included several network structure attributes, such as clustering, centrality, and density. Moreover, they used their framework to present an analysis of two organizations with several hundred employees.

In 2002, Krebs [56] studied Al-Qaeda's organizational network structure attributes following the September 11[th] attacks. They successfully discovered the organization's leader by using the degree and closeness structural properties of vertices.

In 2007, Mishra et al. [57] introduced a new criterion for clustering ubiquitous social networks and provided an algorithm for discovering clusters. They indicated that their algorithm succeeded in finding good clusters.

---

[7] https://twitter.com

## 2.5. Socialbots

We focused on socialbot attacks. In recent years, several adversaries are using them to accomplish many goals [1], [8], [12], [13], [14], [16].

### 2.5.1. Socialbots Definition

Socialbots are defined as automatic or semi-automatic computer programs that control OSN accounts and perform human behaviors such as sending friend requests and messages, and so on [1], [18].

Unlike a regular bot such as a Twitter bot or spambot, a socialbot hides its robot identity. In many cases, socialbots are programmed to infiltrate communities within OSNs and pass themselves off as human beings. Adversaries maintain socialbots for stealing or accessing sensitive information that they are not allowed to access, in the interest of reaching an influential position in order to spread misinformation or propaganda [1], [20].

### 2.5.2. Attacking Socialbots

In recent years, several studies introduced potential socialbot attacks on users.

In 2007, a study was conducted by Sophos regarding the risks of identity and information theft on Facebook [58]. The authors created a fabricated Facebook profile under the name Freddi Staur. Under this name, they manually sent friend requests to two-hundred users across the world. Eventually, eighty-seven out of two-hundred Facebook users accepted the friend request from the fabricated profile, yielding a 41% success rate. Moreover, the authors were able to gather sensitive and personal details about these users because of their acceptance of the requests. The details included their email addresses, dates of birth, phone numbers, photos of family and friends, hobbies, employer, etc.

In 2009, Bonneau et al. [59] investigated the difficulties of collecting profiles and then graphed information from the Facebook OSN. They examined several methods of collecting data—among them, false profiles. They detailed two techniques for working with false profiles. The first was to create false profiles in several networks which could lead to instant access to the profiles of most users. The second technique was to send friend requests to "highly connected users who [were] more likely than average to accept a friend request from a stranger" [59]. They found that creating a false profile and sending friendship requests became effective only with friend-of-friend privacy settings.

In 2009, Bilge et al. [8] demonstrated potential attacks of launching a crawling campaign and identity theft attacks against several OSNs including Xing, StudiVZ,[8] MeinVZ,[9] Facebook and LinkedIn, in order to gather personal and sensitive information. The attacks were divided into two parts: the first attack was a classic identity theft in which Bilge et al. cloned several user profiles of victims and sent friend requests to their contacts. By accepting the friend requests, they were able to access sensitive information of the friends of the victim. Moreover, the second part of the attack was more complex in that the attackers launched a cross-site profile-cloning attack. In this attack, they identified users who had a profile account in one OSN and not in another. They cloned the profile account of the victim in the first OSN and forged a new profile in another OSN where the victim was not registered. Using this method, Bilge et al. rebuilt the OSN of the victim by contacting the victim's friends in both OSNs.

---

[8]http://www.studivz.net
[9] http://www.meinvz.net

Moreover, in 2011, Boshmaf et al. [1] demonstrated the vulnerability of OSNs to a large-scale infiltration campaign by socialbots. They built a network of socialbots and operated them on Facebook for eight weeks. In this campaign, the army of socialbots attempted to connect with many Facebook users. The results included three main conclusions. First, OSNs such as Facebook, can be infiltrated with a success rate of up to eighty percent. They were able to demonstrate that the more friends a user has, the more likely the user is to respond positively to friend requests. Secondly, depending on users' privacy settings, successful infiltration can result in privacy breaches in which even more of the users' data is exposed as compared to that which is exposed with strictly public access. And thirdly, for all practical purposes, the security defense of Facebook, dubbed the Facebook Immune System (FIS), was found not to be very effective in detecting or stopping large-scale infiltration campaigns.

Additionally, Magdon-Ismail et al. [60] studied the infiltration of trust based on an OSN utilizing an agent which sent friend requests. Its mission was to amass as many connections as possible. They developed a model for infiltration based on two properties of actors in the network. First, actors would like to have as many connections with others as possible. Secondly, actors were more likely to connect to trusted nodes. Based on this research, they established a number of conclusions. First, the trust effect is crucial. If an agent is not trusted enough, then it would be difficult to infiltrate a network because of its robustness. Secondly, there is the importance of the network structure. If the trust effect is small, then well-clustered networks like typical OSNs are easier to infiltrate. Moreover, where the trust effect is larger, networks with a large expansion are easier to infiltrate. Thirdly, the algorithm used by the agent is crucial for the success of infiltration. Random requests are less successful with infiltration than greedier strategies.

In 2013, Mitter et al. [61] suggested a categorization scheme of socialbot attacks. Moreover, the recent socialbot attacks are characterized according to the categorization scheme they created.  They defined dimensions for describing and categorizing socialbot attacks: targets, account types, vulnerabilities, attack methods and results.

Also, in 2013, Zhen [15] proposed a spam campaign by using socialbots. In order to do so, he performed several actions, such as creating socialbots, choosing legitimate accounts, forming friendships, earing trust, and spreading spam. His method, which was based on socialbots, earned him the trust of other users in OSNs and allowed him to carry out his spam campaign.

Additionally, Boshmaf et al. [16] continued with their infiltration investigation, generating two novel and significant conclusions. First, they used a mathematical model in order to calculate the profit of an adversary to operate a socialbot network. Eventually, they found that a rational adversary would utilize the socialbot network as a tool for collecting private user information in order to run more profitable, adversarial campaigns, such as email-based spam and phishing campaigns. Secondly, protection against socialbots poses many challenges relating to issues such as web automation, online-offline identity binding, and usable security.

In 2014, Mitter et al. [62] explored the ability of socialbot attacks to influence the social graph on OSNs. They conducted an empirical study in which they tried to understand to what degree socialbots were able to influence link creation between targeted real users within OSNs. In order to test this, Mitter et al. analyzed a dataset from a socialbot experiment on Twitter which was conducted three years before. Eventually, they found that the creation of links within OSNs could not be explained solely by this dataset. Moreover, they suggested that external factors drive link creation behavior.

Moreover, in 2014, Freitas et al. [63] attempted to understand the infiltration strategies of socialbots on the Twitter OSN. They created one hundred and twenty socialbot accounts which differed in their characteristics and strategies. They investigated the extent to which these socialbots were able to infiltrate the Twitter OSN over the duration of a month. Eventually, they found that out of the 120 socialbots they created, only 31% could be detected by Twitter. According to their findings, they suggested that automated strategies were able to infiltrate Twitter defense mechanisms. Moreover, even socialbots which used simple automated techniques to follow other Twitter users and post tweets gained a significant amount of followers and triggered many interactions from other users. Regarding finding the characteristics of socialbots that would enable them to evade Twitter defenses, Freitas et al. found that higher Twitter activity is the most important factor for successful infiltration. Other factors like the gender and the profile image were found to gain some importance.

### 2.5.3. Detecting Socialbots

Along with studies that tried to gather leaked information about users within OSNs and even in some cases to infiltrate them, there were several studies that tried to suggest solutions to these privacy issues. The solutions were based on the detection of fake profiles and socialbots. Quick identification of malicious users within OSNs may help innocent users as well as OSN operators to defend themselves against these malicious users. The techniques for detecting these malicious users were varied and included identification using machine learning [64], [65] as well as creating honeypots for attracting spammers [66].

In 2010, Benevenuto et al. [64] attempted to detect spammers on Twitter. They first collected a large dataset of the Twitter OSN, which included more than fifty-four million users, 1.9 billion links, and almost 1.8 billion tweets. Afterward, they manually classified spammers and non-spammers. They were able to identify several characteristics related to the content of tweets and users' social-behavior for detecting spammers. They used these characteristics for classifying users as either spammers or non-spammers by machine learning techniques. Eventually, they succeeded in detecting approximately seventy percent of spammers and ninety-six percent of non-spammers.

Also in 2010, Chu et al. [33] attempted to help human users in identifying who they were interacting with. They focused on the classification of human, bot and cyborg accounts on Twitter. In order to detect them, they crawled the Twitter OSN and collected information regarding half a million accounts on Twitter. Eventually, they found differences between human, bots and cyborgs in terms of tweeting behavior, tweet content, and account attributes. Based on their findings, they suggested an automated classification system. This system included four components: an entropy-based component, a machine-learning component, an account properties component, and a decision maker. This system took into account all of these components and determined the likelihood of a given user to be a human, bot, or cyborg.

Likewise in 2010, Lee et al. [66] proposed a honeypot-based approach for discovering OSN spammers. Their approach was based on creating social honeypots within the MySpace and Twitter OSNs in order to attract spammers to attack. With these honeypots, they developed statistical user models in order to distinguish between social spammers and legitimate users. Eventually, their honeypots succeeded in identifying social spammers with low false positive rates.

In 2010, Stringhini et al. [67] created several "honey-profiles" on three large OSNs: Facebook, MySpace, and Twitter. Later, they analyzed the collected data and were able to identify anomalous behaviors of users. Eventually, they were able to detect and delete 15,857 spam profiles. Through these actions, they succeeded in proving that it is possible to automatically identify the profiles that were used by spammers.

In 2011, Ratkiewicz et al. [42] detailed the abuse of microblogging services like Twitter for illegitimate use. Their study focused on political astroturf which is defined as a campaign disguised as spontaneous, popular behavior, when in reality these campaigns are carried out by a single person or organization. Ratkiewicz et al. demonstrated a web service that follows political memes on Twitter. Memes are defined as elements of a culture that may be considered to be passed by non-genetic means, especially imitation [68]. The researchers used this web service in order to detect astroturfing, smear campaigns and other misinformation in the context of U.S.

elections. In this study, the researchers emphasized the importance of the early identification of accounts associated with astroturf memes.

In 2012, Fire et al. [65] developed an algorithm for the detection of spammers and fake profiles in OSNs. Their method was based solely on the topological features of the OSN. Fire et al. attempted to detect anomalies in the topology of OSNs. The proposed method is based on a combination of graph theory algorithms and machine learning and has been evaluated based on the datasets of three different OSNs: Academia.edu,[10] AnyBeat,[11] and Google+.[12] The dataset from Academia.edu contained more than 200,000 users and almost 1.4 million links. Regarding AnyBeat, the topology contained 12,645 users and 67,053 links, and the dataset from Google+ contained more than 211,187 users and 1,506,896 links. Eventually, the researchers were able to detect several fake profiles. Furthermore, they found differences regarding the characteristics of each of the OSNs and their users. They concluded that their algorithms could provide sufficient protection for small and medium sized OSNs with several million users. However, the algorithms were not sufficient for large scale OSNs.

In 2013, Wang et al. [69] built a system for the identification of fake profiles using server-side clickstream models. The detection process was based on converting similar user clickstreams into behavioral clusters by capturing distances between clickstream sequences. Wang et al. tested their developed models on 16,000 real and Sybil users from Renren,[13] a large Chinese OSN. Eventually, they showed that their method worked with high detection accuracy.

Also in 2013, Wald et al. [70] explored a dataset that contained 610 users, who received messages from Twitter bots. They attempted to find features that helped in detecting bots. After, they built models with classifiers for predicting if a user would interact with a bot.

In 2014, Wagner et al. [18] thought that modern social media security defenses needed to advance in order to be able to detect social-bot attacks. In their opinion, identifying socialbots was crucial. As a result, identifying users who were susceptible to socialbot attacks - and implementing the means to protect against such attacks - was also important. Wagner et al. defined two separate identities: *target* and *susceptible user*. A *target* represents a user who has been chosen by socialbots to be a target for an attack. A *susceptible user* is a user who has been infected by a socialbot. This kind of infection is defined as cooperation of the user with the agenda of a socialbot. They attempted to identify factors for detecting users who were susceptible to socialbot attacks. In order to find these factors, Wagner et al. studied data from an experiment conducted by WebEcologyProject. The experiment was termed, the Social Bot Challenge 2011. In this experiment, three teams implemented several socialbots. Each one of them attempted to influence user behavior on the Twitter OSN. The

---

socialbots persuaded targets to interact with them by replying to them, mentioning them in their tweets, retweeting them or following them. The group of targets included 500 unsuspecting Twitter users which were selected based on the fact that they had an interest in or tweeted about cats. A susceptible user was defined as a target that interacted at least once with a socialbot. For example, the user replied, mentioned, retweeted or followed the socialbot. They used the data in order to develop models for identifying susceptible users among a given set of targets and predicting users' level of susceptibility. Eventually, they were able to introduce three groups of features: network, behavior and linguistic features.

In 2014, Paradise et al. [17] investigated monitoring strategies for detecting socialbots within OSNs. They evaluated socialbot attack and defense strategies using a simulation tool. The simulation consisted of nodes and connections on real OSN data. They implemented four attack strategies and six defense methods.

Recently, Abulaish et al. [71] developed an ensemble learning method for detecting spammers. They evaluated the performance of several basic ensemble classifiers over individual features of legitimate users and spammers on OSNs. In their study, Abulaish et al. extracted structural features from user interaction patterns of OSN users. Their real dataset included wall post activities of 63,891 Facebook users. Finally, they found that the bagging ensemble learning approach using the J48 decision tree performed better than an individual model, and also than other ensemble learning methods.

### 2.5.4. Security Tools for Protecting Users

In recent years, spammers, and hackers have found OSNs to be an efficient platform for spreading malware and spam. Moreover, Rahman et al. [72] found that at least thirteen percent of the applications on Facebook are malicious. The development of designated tools for users' protection is one of the ways to protect users on OSNs against spammers and hackers.

For example, in 2011, Stein et al. [46] described the Facebook Immune System (FIS). Like the human immune system, the FIS is described as a system, which in this case, protects Facebook from adversaries and malicious attacks. The FIS is an adversarial learning system which performs real-time checks and classifications on every read-and-write action in Facebook's database. These checks exist in order to protect Facebook users and the entire social network from malicious activities. Furthermore, Stein et al. described the design of the FIS and the challenges faced by the FIS.

Due to the significant problem of hackers' usage of applications for the potential spreading of malware and spam, in 2012, Rahman et al. [72] developed the FRAppE tool for detecting malicious Facebook applications. They found a set of attributes that helped to distinguish between malicious applications and benign ones. Moreover, they revealed that FRAppE can detect malicious apps with 99.5% accuracy, with no false positives and a low false negative rate (4.1%).

In 2013, Fire et al. [73] developed the Social Privacy Protector (SPP) software for improving users' security and privacy on Facebook. The Social Privacy Protector software contained three protection layers. The first layer analyzed Facebook users' friend lists in order to recognize which friends of users were thought to be fake. This application enabled the users to restrict access to the users' personal information via fake profiles. The second layer expanded Facebook's basic privacy settings based on different types of OSN user profiles. The last layer warned users about the number of installed applications on their Facebook profile. More than 3,000 users from more than twenty different countries installed this software, with 527 actually using the software. Overall, users who installed the SPP restricted more than nine thousand users, and at least 1,792 Facebook applications were removed. By analyzing the dataset obtained by this software, together with machine learning techniques, they developed classifiers in order to predict which Facebook profiles had a high probability of being fake, therefore, posing a risk to users.

Furthermore, in 2013, Kagan et al. [74] developed a Firefox add-on that alerted users to the number of installed applications on their Facebook profiles. The dataset they collected consists of data from 2,945 users. Eventually, they found a linear correlation between the average percentage change of newly installed Facebook applications and the number of days passed since the user initially installed their Firefox add-on. Furthermore, they discovered that users who used the add-on became more aware of security and privacy issues as this was reflected by the fact that these users installed, on average, fewer new applications.

# 3. Detailed Description of the Conducted Research

## 3.1.     Research Hypotheses

In recent years, adversaries have found OSNs to be favorite targets for attacks. Most of them have chosen to carry out their attacks using socialbots [62]. These malicious users use socialbots for several reasons. Some of the adversaries use socialbots for influencing users or spreading misinformation and propaganda. One famous manifestation of spreading misinformation and propaganda by socialbots was during the U.S. political elections [41], [62]. Others apply socialbots for reconnaissance regarding organizations and employees [17]. Another group of malicious users use socialbots in order to infiltrate users [1], [8], [12], [13], [14]. All kinds of these attacks could result in fraud as well as loss of intellectual assets and confidential business information [11].

Our hypothesis was divided into two assumptions. The first assumption was based on the fact that adversaries can utilize socialbots for mining the data of targeted organizations. In this study, we used two types of socialbots: passive and active socialbots.

Passive socialbots are defined as public user accounts that we created in order to crawl and gain employees who worked or currently work in a given targeted organization. These passive socialbots were created only for crawling purposes. These socialbots do not send friend requests to other OSN users. Therefore, they have no friends within an OSN.  By crawling with passive socialbots, we expected to gather the public information regarding a targeted organization, which includes organizational connections and employees.

As opposed to passive socialbots, active socialbots are defined as public user accounts that perform various types of social activities, such as uploading posts, or sending friend requests to other OSN users. By using active socialbots, we attempted to find new additional organizational employees and inner organizational connections within OSNs that we could not find by crawling with passive socialbots. These potential employees could be employees who have more strict privacy settings than those who we would be able to find with the passive socialbots.

Furthermore, our second assumption was based on previous studies, which noted that many OSN users are unaware of the privacy issues that accompany the use of OSNs [10]. We validated this assumption by using socialbots. We attempted to understand to what extent users are unaware of various privacy issues. This validation process was carried out in hopes that our socialbot would be accepted as a friend by specific employees in targeted organizations.

These two assumptions were tested and are described in the next sections. Both assumptions emphasize the privacy issues which exist within OSNs.

## 3.2.     Methods

In this section we present two algorithms which involve the use of socialbots. The first algorithm is considered organization data mining, whereas the second algorithm is considered reaching specific users in targeted organizations.

Most OSNs require users to create accounts in order to establish connections with other network users [61]. In many cases, a user's account may include personal data such as photographs, birthday, hometown, ethnicity, and personal interests [1]. In most undirected OSNs, users connect to one another by transmitting friend requests. The recipient must accept the friend request in order to establish a friend link with the initiator of the request. When this process has been completed, the two parties acquire the privilege of accessing each other's profile details at will. Therefore, we decided to define an accepted friend request as an infiltration to the social network of the user.

Our algorithms were based on the establishment of socialbots on OSNs. The idea was that when users accept these socialbots' friend requests, these socialbots gain increased access to users' profiles and may gather additional information about users and, in some cases, information about the user's friends [1], [8], [13], [14], [16].

The following sections (see Section 3.2.1 and 3.2.2) present the two algorithms for testing our assumptions thoroughly. Section 3.2.1 focuses on the first assumption while Section 3.2.2 focuses on the second.

### 3.2.1.  The Mining of Data of a Targeted Organization

To mine information regarding a targeted organization, several actions, as depicted in *Algorithm* 1, had to take place. First, we had to crawl the targeted organization's website and gather public information about its employees, who have a Facebook user account. For the crawling process, we created a passive socialbot (referred to as P) on Facebook, with no friends. We used the Organization Social Network Crawler, developed by Fire et al. [11]  and crawled with P (see line 1). This special crawler optimizes data collection from users associated with a specific group or organization. It distinguished from other standard crawlers by utilizing the homophily principle [75], which states that the more specific person's friends have been employed by an organization, the more likely that this person was employed by the same organization at some point.

After finishing crawling and gathering intelligence on the targeted organization's employees with the passive socialbot P, we created an active Facebook socialbot account (referred to as S) for every organization that we wanted to reach (see line 2).

Before reaching a targeted organization, it was essential that the socialbot account profile look like a reliable profile of an actual person. For this reason, we added personal attributes of real profiles such as posts, images, interests, etc. [1], [8] (see line 3)

Initially, we manually sent friend requests to fifty random users regardless of any organizational affiliation, who had more than a predefined threshold—such as a minimum of a thousand friends (see line 4 -7 ). The rationale for doing this was again based on Boshmaf et al.'s [1] observations which stated that "the more friends [a user] had, the higher the chance was that they [would] accept a friend request from a socialbot (i.e., a stranger)."

After fifty random users accepted our socialbot friend requests, we started to automatically send friend requests to employees working in a targeted organization. In order to choose the right employees to send friend requests to, we sent friend requests to the employees of the target organization who had the highest number of friends (see line 8 - 10). It is important to mention that we conducted preliminarily intelligence manually, which helped us to reconstruct parts of the organization's social network, which revealed the users with the highest number of Facebook friends.

After a socialbot made friends with at least ten organizational employees, we sent friend requests to the employees with the highest number of mutual friends with our socialbot in descending order (see line 11 - 13). We finished sending friend requests when we did not have employees that had a single mutual friend with our socialbot.

---

### Algorithm 1: Socialbot organizational intrusion

**Input:** *Uids* - A set of seed URLs to Facebook profile pages of
       organization's employees, $P$ – Passive socialboat, $S$ – Active socialboat,
       $O$ – targeted organization
**Output:** A set of Facebook profiles and their connections
1:    *OrgPublicGraph* ← Organizational-Crawler($P$, *Uids*, $O$)
2:    $S$ ← CreateActiveSocialbot($S$, $O$)
3:    $S$ ← CreateReliableProfile($S$)
4:    $i$ ← 0
5:    **while** (NumOfFriends($S$) <= 50)
6:           SendFriendRequestToRandomUsers($S$)
7:    **end while**
8:    **while** (NumOfOrgFriends($S$) <= 10)
9:           SendFriendRequestToOrgEmpMaxNumOfFriends($S$)
10:   **end while**
11:   **while** (NumOfOrgFriends($S$) <= MaxNumOfFriends)
12:          SendFriendRequestToOrgEmpMaxMutualFriends($S$)
13:   **end while**
14:   **return** Collect pages

---

### 3.2.2. Reaching Specific Users in Targeted Organizations

The suggested algorithm involved several actions: First, as it was depicted in *Algorithm 1* we had to crawl targeted organizations and gather public information about employees who established user accounts and stated that they were working or worked in the targeted organizations in the past. The crawling process was based on the Facebook, and Xing OSNs and a passive socialbot account, P, that we created. The crawler was similar in its behavior to that which was introduced by Fire et al. [11]. The crawler received, as an input, a list of user IDs who were employees in the targeted organization. It crawled their friends list and was able to gather users who were employees in the targeted organization (see line 1).

Second, similarly to *Algorithm 1*, and prior to the infiltration of a targeted organization, we designed a user profile for our socialbot. It was important that it seemed like a reliable profile of an actual user, e.g., including photos, posts, etc. (see line 2 - 3).

Third, by the end of the crawling process, we gained sufficient information about users who were working or had worked in the targeted organizations and their connections. In other words, we gathered intelligence on the targeted organizations' employees, which we utilized in order to randomly select ten users to serve as targets for the reaching process. We then utilized their mutual friends in order to reach these targeted users (see line 4 - 8).

Afterward, like in *Algorithm 1*, the active socialbots suggested friend requests to fifty random users that had more than one thousand friends (see line 9 - 11).

After a socialbot succeeded in gaining a positive response to its friend request from fifty random users, it automatically sent friend requests to targeted users' mutual friends who were employees in the same organization (see line 12 - 15). We then waited for the socialbot to gain fifty friends from random users as we wanted our socialbots to appear as much like genuine users as possible. We did not send friend requests to mutual friends of the targeted organizations when our socialbot had only a limited number of friends, as users tend not to accept requests from other users with only a few friends when they initiate a friend request. A lack of friends could cause other users to automatically reject our socialbot's friend requests [12].

Fourth, for each chosen targeted user, we detected his or her friends inside the organization, and our socialbot sent friend requests to them. The process of sending friend requests was handled in descending order based on the targeted user's number of friends: at first the socialbot sent friend requests to the most "friendly" users, i.e., those with the largest number of friends, and eventually, requests were sent to users with the fewest friends in the targeted organization. The idea was to gain as many mutual friends as possible, an accomplishment that may have increased the probability that a socialbot's friend request would be accepted by a targeted user.

Fifth, after the completion of the process of sending friend requests to targeted users' mutual friends, we sent friend requests to the ten targeted users. Afterward, we observed how many of them accepted our socialbot's friend requests (see lines 16 - 18).

## Algorithm 2: Socialbot specific intrusion

**Input:** *Uids* - A set of seed URLs to Facebook profile pages of an organization's employees,
  *P* – Passive socialbot,
  *S* – Active socialbot,
  *O* – targeted organization,
  TU – targeted users,
  OG – organization's graph
*1: OG* ← Organizational-Crawler(*P*, *Uids*, *O*)
2: *S* ← CreateActiveSocialbot(*S*, *O*)
3: *S* ← CreateReliableProfile(*S*)
*4: i* ← 0
5: **while** ($i < 10$) **do**
*6:*      *TU* ← RandomizeTargetedUsers(*O*)
*7:*      *i* ← *i* + 1
8: **end while**
9: **while** (NumOfFriends(*S*) <= 50)
10:    SendFriendRequestToRandomUsers(*S*)
11: **end while**
*12: TUFriends* ← FindOrgFriends(*O*, *TU*, *OrgPublicGraph*)
13: **for** ( f ∈ *TUFriends*)
14:    SendFriendRequestInDescendingOrder (*f*)
15: **end for**
16: **for** (*U* ∈ *TU*)
17:    SendFriendRequest (*U*)
18: **end for**

### 3.3.      Targeted Online Social Networks

In this section, we provide a brief overview of each one of these networks. Additionally, we describe in detail each of the organizations we crawled and analyze the datasets we were able to gather.

#### 3.3.1. Datasets

In this section we present the datasets we worked on. We carried out *Algorithm* 1 (see Section 3.2.1) on Facebook. In regards to *Algorithm* 2 (see Section 3.2.2), we tested it on two different OSNs: Facebook and Xing.

##### 3.3.1.1. The Mining of Data of Targeted Organizations

In this section we present the datasets on which we tested *Algorithm* 1. All the targeted organizations were gathered from the Facebook OSN.

**Targeted Organizations.** We decided to conduct our data mining on three organizations: $O_{F1}$, $O_{F2}$, and $O_{F3}$.

The $O_{F1}$ organization is an international software company that develops, licenses, implements, and supports software applications for its customers. According to public sources, the $O_{F1}$ organization employs thousands and operates offices in North America, Europe, and the Middle East.

In the primary crawling process with the passive socialbot P, we were able to expose 1,859 informal connections of 309 Facebook users who noted on their Facebook profiles that they were employees in this targeted organization (see Table 1, and Figure 1a).

| Organizations | Nodes | Edges | Clusters | Cluster Average Size | Cluster Max Size |
|---|---|---|---|---|---|
| $O_{F1}$ | 309 | 1,859 | 35 | 8.6 | 92 |
| $O_{F2}$ | 413 | 3,536 | 19 | 21.421 | 130 |
| $O_{F3}$ | 1,484 | 19,484 | 141 | 9.851 | 268 |

Table 1: Organizational datasets statistics

Moreover, by using the Markov Clustering (MCL) algorithm [76] we were able to discover 35 clusters with an average size of 8.6 employees, and a maximal size of 92 employees (see Figure 1b). MCL is a way to cluster a graph by flowing through a network. This algorithm has been widely used for clustering in biological networks. However, it demands that the graph be sparse, and therefore we found it problematic at times for our purposes. However, we tested many other clusters, and this was the only algorithm that found a different number of clusters when we compared the datasets P provided by crawling and the datasets the socialbot provided.

The $O_{F2}$ organization is a leading information technology company that specializes in the integration, development, and application of technologies, solutions and software products, hardware, infrastructure, etc. The company is located in Eastern Europe and the Middle East and has thousands of employees around the world.

Primarily, in the crawling process with the passive socialbot P, we were able to expose 3,536 informal connections of 413 Facebook users, who listed on their Facebook profiles that they were employed by this targeted organization (see Table 1, and Figure 2a). Furthermore, by using the MCL algorithm we were able to discover 19 clusters with an average size of 21.421 employees and a maximal size of 130 employees (see Figure 2b).

The $O_{F3}$ organization is a technology company which develops, and markets telecommunications software. Moreover, this organization focuses on providing services to telecommunication service providers. According to public sources, the $O_{F3}$ organization was founded in the 1980s and employs approximately 5,000 employees in the United States of America and the Middle East.

In the primary crawling process with the passive socialbot P, we were able to expose 19,484 informal connections of 1,484 Facebook users, who stated on their Facebook profiles that they were employees in this targeted organization (see Table 1, and Figure 3a). Furthermore, by using the MCL algorithm we were able to discover 141 clusters with an average size of 9.851 employees and a maximum size of 268 employees and minimum size of 2 employees (see Figure 3b).

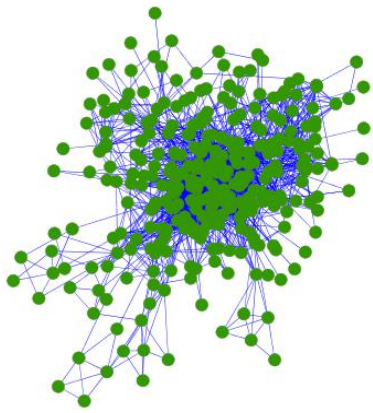

**(a)** Organizational social
network has been crawled
by P

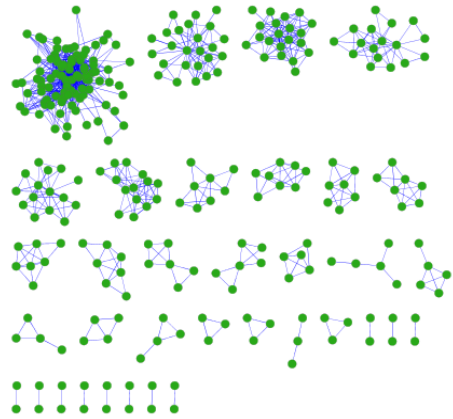

**(b)** Clusters (before intrusion)

**Figure 1: OF1 organization**

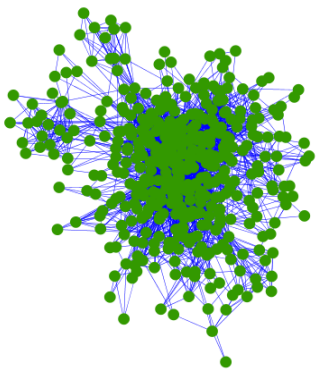

**(b)** Organizational social
network has been
crawled by P

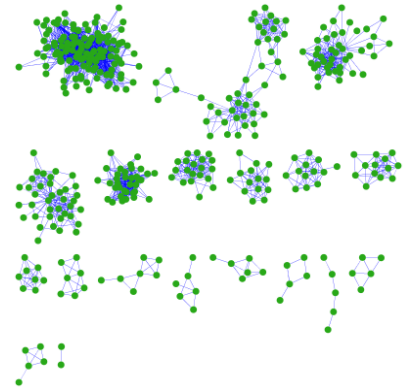

**(a)** Clusters (before intrusion)

**Figure 2: OF2 organization**

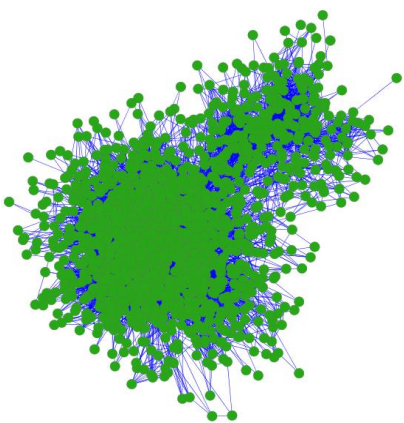

**(a)** Organizational social network
has been crawled by P

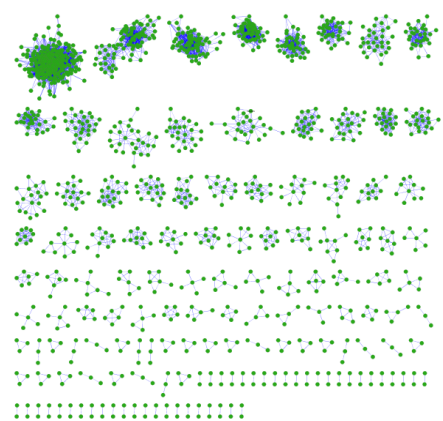

**(b)** Clusters (before intrusion)

**Figure 3: OF3 organization**

### 3.3.1.2. Reaching Specific Users

We decided to reach specific employees who stated on their Facebook profile that they were employees of one of four targeted organizations. Eventually, we succeeded in reaching three of these organizations: $O_{F4}$, $O_{F5}$, and $O_{F6}$.

Furthermore, on Xing, we used five socialbots for specific user infiltration of five different organizations. Eventually, we succeeded in infiltrating two of the organizations. We present the results from the attempts of $S_{X1}$, $S_{X2}$, $S_{X3}$, $S_{X4}$, and $S_{X5}$ socialbots that tried to reach specific employees in the targeted organizations $O_{X1}$, $O_{X2}$, and $O_{X3}$ respectively (see Section 4). $S_{X4}$ and $S_{X5}$ were blocked in the first stage so they did not attack an organization.

**Targeted Organizations**. The $O_{F4}$ organization is the same $O_{F1}$ organization that is mentioned in Section 3.3.3.1.[14]

In the crawling process we identified 2,199 informal links of 330 Facebook users who, according to their Facebook profiles, work or have worked in this organization (see Table 2 and Figure 4a).

| Organizations | Social Networks | Nodes | Edges |
|---|---|---|---|
| $O_{F4}$ | Facebook | 330 | 2,199 |
| $O_{F5}$ | Facebook | 469 | 3,831 |
| $O_{F6}$ | Facebook | 918 | 10,986 |
| $O_{X1}$ | Xing | 107 | 369 |
| $O_{X2}$ | Xing | 1,237 | 14,408 |
| $O_{X3}$ | Xing | 416 | 4,153 |

Table 2: Organizational datasets statistics

The $O_{F5}$ organization is the same $O_{F2}$ organization that is mentioned in Section 3.3.3.1.
In the crawling process we discovered 3,831 informal links of 469 Facebook users who, according to their Facebook profiles, work or have worked in this organization (see Table 2 and Figure 4b).

The $O_{F6}$ organization is a telecommunications networking product provider that provides communication products and develops services for carriers, cable and multiple system operators, wireless/cellular service providers, etc. The $O_{F6}$ organization is located in the Middle and Far East, and according to public sources, maintains thousands of employees. In the crawling process, we identified 10,986 informal links of 918 Facebook users who, according to their Facebook profiles, work or have worked in this organization (see Table 2 and Figure 4c).

The $O_{X1}$ organization is a large crude oil and natural gas producer. The activities of this organization include exploration and production of oil and natural gas as well as the natural gas trade, including transport and storage. This organization is based in Eastern and Western Europe, North Africa, and South America. $O_{X1}$ maintains thousands of employees worldwide. In the crawling process, we identified 369 informal links of 107 Xing users who, according to their profiles, work or have worked in this organization (see Table 2 and Figure 4d).

---

[14] In order to differentiate between experiments, we used different names for the organizations.

The $O_{X2}$ organization is a food company that serves as a manufacturer, retailer, and marketer of beverage concentrates. The organization is situated in the U.S but has branches worldwide. It manages hundreds of thousands of employees worldwide, and in this study we focused on the European branch of the $O_{X2}$ organization. In the crawling process we identified 14,408 informal links of 1,237 Xing users who, according to their profiles, work or have worked in this organization (see Table 2, and Figure 4e).

The $O_{X3}$ organization is a corporation that develops, manufactures, and sells computer software, electronics, etc. This organization is American, but has numerous branches worldwide. $O_{X3}$ has hundreds of thousands of employees worldwide. In this study, we focused on the European branch. In the crawling process we identified 4,153 informal links of 416 Xing users who, according to their Facebook profiles, work or have worked in this organization (see Table 2, and Figure 4f).

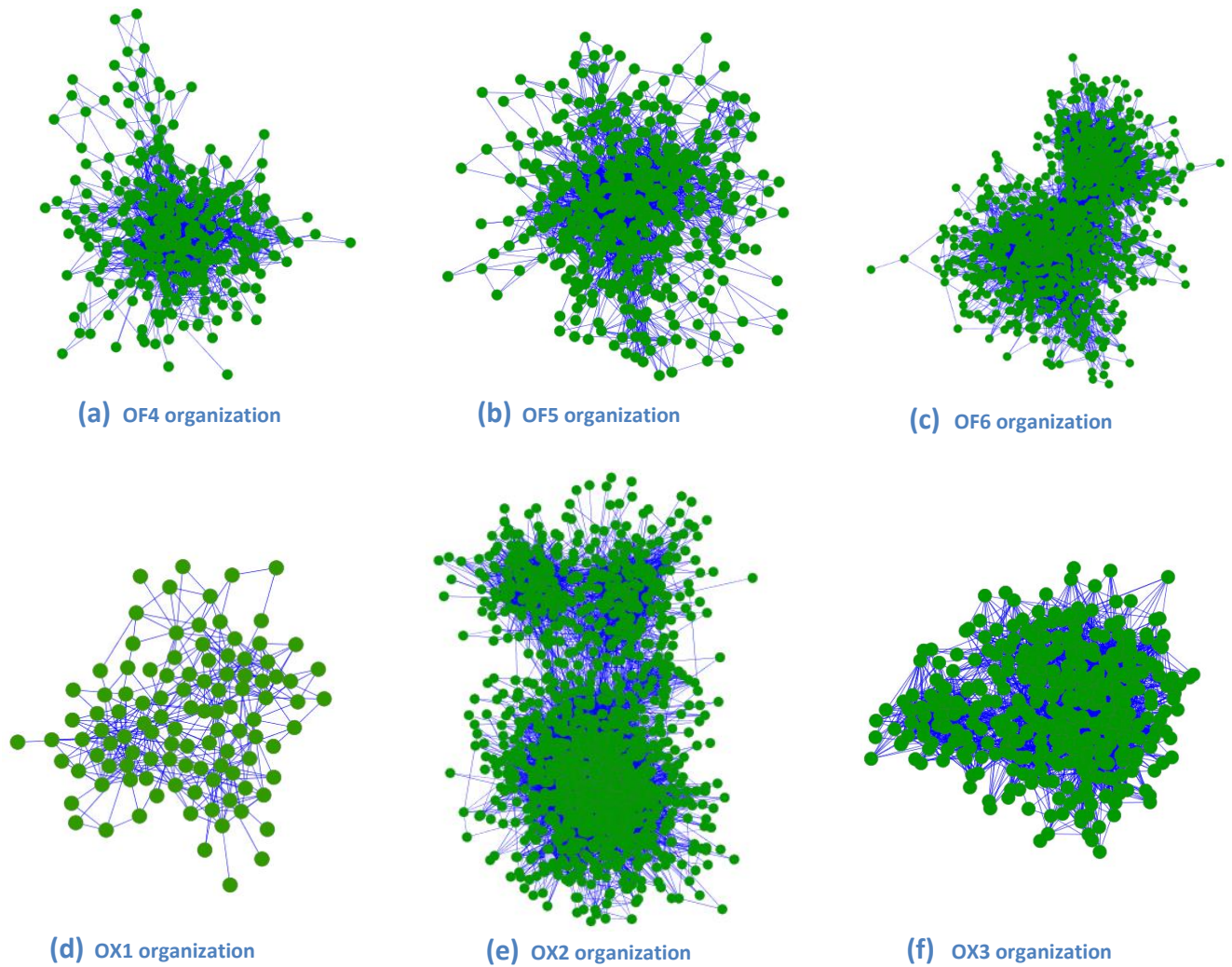

**(a)** OF4 organization    **(b)** OF5 organization    **(c)** OF6 organization

**(d)** OX1 organization    **(e)** OX2 organization    **(f)** OX3 organization

Figure 4: Organizational social network

## 3.4.    Prevailing Difficulties

The process of mining data regarding targeted organizations as well as reaching specific users employed in targeted organizations revealed several obstacles. These complicated processes contain several sub-processes including crawling the targeted organization's OSN, creating fabricated profiles for the identity of our socialbots, and finally reaching specific employees of the organizations. Each of these tasks presented challenges. The operators of OSNs should consider the challenges associated with these processes and work on preventing them.

### 3.4.1.  Adjusting the Crawling Process

The crawling phase is a fundamental step in the general processes. This phase includes several sequential actions: First, we had to select an organization as a target. The selected organization could be any organization or company with employees who use the Facebook or Xing OSNs. On Xing, it is much easier to find employees by a given organization because it is designated for professionals, who in most cases, display their workplace's name. We could also make use of existing pages that were created and operated by organizations in order to present themselves and display their activities publicly. In many cases, these pages included minimal information (for example, a small list of employees), that served as an information source for the crawling process.

After we chose our targeted organizations, we had to create a passive socialbot to initiate crawling on the OSNs. With these public accounts, we did not try to reach other employees, but rather to crawl on the OSNs and find user profiles that fit the given criteria. These accounts did not include social properties except for a name and an image of an animal of some kind. Moreover, they did not send any friend requests to anyone.

When we finished creating a passive socialbot account, and after we had already chosen a targeted organization, we had to connect to Facebook or Xing with this created passive socialbot account and operate the crawler. As the crawler ran, it provided us with the profiles as text or HTML files, and also provided a list of connections between the users. The full implementation is described by Fire et al. [11].

In the described procedure, the crawler was able to collect between hundreds and thousands of OSN user profiles who defined themselves as employees in the given targeted organization within a few days, depending on the targeted organization and its size. Actually, we could have gained many more user profiles, but we added time-outs and delays for slowing the crawler's actions: a timeout of $0.5 - 1$ minute for downloading profile, and a delay of 8 hours between crawling processes. It is not recommended to download a large amount of profiles over a short time. The OSN providers may identify the abnormal behavior of the crawler as malicious because it is uncommon for a user to surf hundreds of profiles so rapidly. One more option would be that if a great number of users would behave like the crawler, it would hurt the performance of the OSN server. Arousing the suspicion of the OSN providers could result in blocking the passive socialbots for crawling.

### 3.4.2. Creating Realistic Socialbot Profiles

The process of creating socialbot accounts within Facebook and Xing was largely done manually. The challenge faced at this stage was to avoid suspicion and to appear as a regular OSN user; otherwise our socialbots' friend requests would have failed. In order to prevent failure, we chose common names for our socialbot accounts with the intent of looking familiar to other users. Then, we had to select images for socialbots' profiles.

In contrast to Boshmaf et al. [1], for female users on Facebook we selected obscure images of real women, such as images without a face, in order to make recognition unfeasible. For male users, we chose images of puppies, fancy cars, etc.

To look like an authentic user, we added interests and other properties for each socialbot. The properties were "likes" to popular singers and movies, posting high quality nature images, adding posts to the user's timeline, etc.

We based the selection of profile images for our Xing socialbots on a unique profile image which included a suit and a tie in order to look like a professional employee. The image was based on profile images of two or three real people, which we then combined to create a new image of a person who does not actually exist.

### 3.4.3. Reaching Specific Employees

In the next stage, our socialbots were to create new connections with fifty users. These users would be picked randomly; however, each chosen user should have more than a thousand friends. Eventually, this stage passed without any specific problems on Facebook. Users with more than a thousand friends tended to accept strangers' friendship requests, and the high percentage of friends we gained helped our Facebook socialbots look like real users to other potential friends. However, because of the small size of the Xing OSN compared to Facebook, it was hard for us to find fifty random users, who had more than a thousand connections. In order to overcome this obstacle, we reduced the threshold of the number of friends a potential user needed from a thousand to four hundred on Xing.

Moreover, one of the parameters that can be used to detect suspicious activity is the community structure of users [60]. Therefore, socialbots are allowed to send friend requests to users only in a limited number of such communities. OSN providers use algorithms for detecting anomalies in a suspicious account's network topology. According to Bosmaf et al. [1], users who connect randomly with strangers may actually be fake profiles. Hence, most of the legitimate users are connected only to a small number of communities. Fake profiles, on the other hand, tend to establish friendships with users from a large number of different communities [65]. In order to overcome these obstacles, we sent few friend requests in the beginning of the process to users who had a large number of friends. After a "friendly user" accepted the socialbot's friend request, our next friend requests were sent to mutual friends of the "friendly user"—who also had a large number of friends.

Afterwards, in the next phase we had to face numerous challenges. The second phase involved sending friend requests to mutual friends of a targeted user in the targeted organization. In this phase, a socialbot may be blocked or disabled by the OSN. The process of disabling users is based on the amount of users that accept (or decline) a socialbot's friend requests. A low acceptance percentage on Facebook, as well as on Xing, can trigger the OSN's anomaly detection mechanism. As a result, a socialbot can get a warning regarding misuse suspicion, i.e., the OSN provider may accuse a socialbot of not knowing the specific user to whom he or she sent a friend request. Moreover, in exceptional cases, Facebook providers may send a message indicating that they have decided to halt the friend request in order to prevent misuse. If a socialbot continues to send friend requests while a low rate of acceptance persists, Facebook may then block this socialbot. Once blocked, Facebook requires a socialbot to provide a real phone number to verify its identity in order to avoid being disabled or losing access to Facebook.

Furthermore, Facebook provides different defense mechanisms like checking the IP address of the logged-in user, facial recognition, and so on. We had to take these matters into account while conducting our study, in the interest of not disturbing the regulatory activity of the OSN and not to bias the study as well.

On Xing, we did not receive warnings prior to being blocked. In the case of blocking a user, a user logs on to Xing and simply receives a message that the account has been deactivated. In some cases, our socialbots were blocked by Xing providers probably as a measure of protection. Xing providers count the number of unconfirmed users, i.e., users who did not confirm a specific user. When the count reaches one hundred unconfirmed users, Xing prevents the user from sending any additional friend requests. In order to overcome this obstacle we made an earnest attempt to perform the minimum amount of activities possible on Xing.

# 4. Results

Two experiments were conducted in this research. One experiment tested *Algorithm 1* (see Section 3.2.1), whereas the second experiment tested *Algorithm 2* (see Section 3.2.2). The following results are presented in the following two sections.

## 4.1.     The Mining of Data of a Targeted Organization[15]

We used three active socialbots on Facebook for mining information from three different organizations. First, the socialbot named $S_{F1}$ for reaching the $O_{F1}$ organization, the second socialbot named $S_{F2}$ for reaching the $O_{F2}$ organization, and finally, the socialbot named $S_{F3}$ for reaching the $O_{F3}$ organization. Moreover, we operated a passive socialbot account, referred to as P. P was a public Facebook user account without friends. We used P only for crawling as an anchor point. P had no friends, so crawling with P was able to provide publicly available data.

After we finished the crawling process using P, we utilized the $S_{F1}$, $S_{F2}$ and $S_{F3}$ active socialbots for reaching the $O_{F1}$, $O_{F2}$, and $O_{F3}$ organizations, respectively, using the techniques described in *Algorithm 1* (see Section 3.2.1). The intrusion process by our socialbots was completed successfully when we crawled again on the targeted organizations, but this time with $S_{F1}$, $S_{F2}$ and $S_{F3}$ instead of P. As we can observe in Table 3, the $S_{F1}$, $S_{F2}$, and $S_{F3}$ socialbots gained 142 mutual friends of the targeted users in the targeted organizations so they were able to uncover more hidden connections and users in the targeted organizations.

In the first process, $S_{F1}$ was able to gain 57 Facebook users who declared in their profile that they were currently $O_{F1}$ employees or past $O_{F1}$ employees. Overall, $S_{F1}$ sent 126 friend requests to 126 different $O_{F1}$ employees and the rate of acceptance was 45.24% (see Table 3a, and Figure 5). $S_{F2}$ gained 60 $O_{F2}$ employees, while $S_{F2}$ sent 107 friend requests to $O_{F2}$ employees, which indicates a rate of acceptance of 56.07% (see Table 3b, and Figure 5). As opposed to $S_{F1}$ and $S_{F2}$, $S_{F3}$ achieved 25 $O_{F3}$ employees that became its friends. $S_{F3}$ sent 56 friend requests to 56 different $O_{F3}$ employees and achieved a rate of acceptance of 44.64% (see Table 3c, and Figure 5).

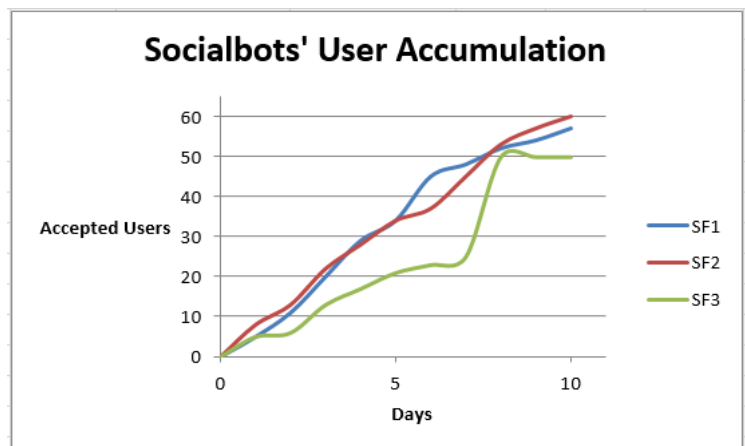

**Figure 5: Socialbots' user accumulation**

---

[15] The results shown in this section are also published in the paper "Organizational Intrusion: Organization Mining using Socialbots," Section IV.

Next, we crawled with $S_{F1}$ on the $O_{F1}$ organization in order to uncover new employees and new connections that we did not find when we crawled with P. Finally, we identified 2,199 informal connections of 330 Facebook users who, according to their Facebook profiles, worked at the $O_{F1}$ organization (see Figure 6a). This means that $S_{F1}$ discovered 6.79% more employees, and 18.29% additional hidden connections (see Table 4). By crawling with $S_{F2}$, we were able to identify 3,831 informal connections of the 469 $O_{F2}$ employees (see Figure 6b).

This indicates that $S_{F2}$ was able to uncover 13.56% more employees, and 8.34% more hidden links (see Table 4). Moreover, by crawling with $S_{F3}$, we detected 23,823 informal connections of the 1,675 $O_{F3}$ employees (see Figure 6c).

This means that $S_{F3}$ detected 12.87% more employees, and 22.27% more hidden links (see Table 4).

Moreover, using the MCL algorithm, $S_{F1}$ was able to discover 34 clusters with an average size of 9.5 employees, and a maximal size of 159 employees (see Figure 7a). $S_{F2}$ was able to discover 29 clusters with an average size of 15.828 employees, and a maximal size of 257 employees (see Figure 7b), and $S_{F3}$ was able to discover 163 clusters with an average size of 9.632 employees, and a maximal size of 296 employees (see Figure 7c).

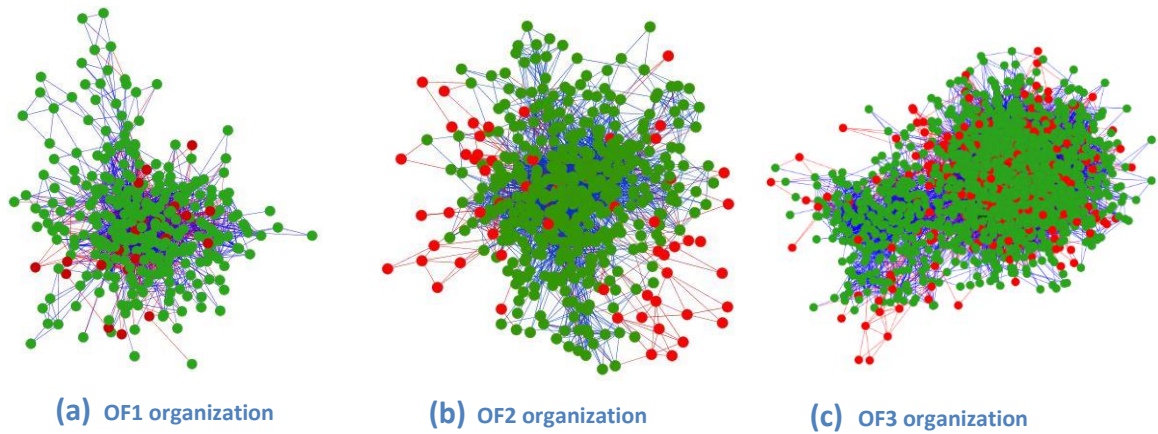

**(a)** OF1 organization          **(b)** OF2 organization          **(c)** OF3 organization

**Figure 6: Organizational social network crawled by active socialbot (red links represent newly discovered links, and red nodes represent newly discovered employees)**

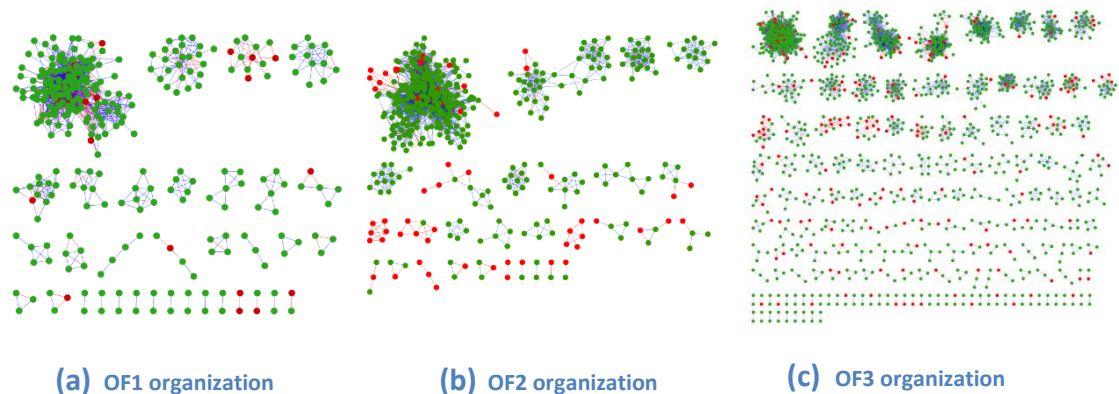

**(a)** OF1 organization          **(b)** OF2 organization          **(c)** OF3 organization

**Figure 7: Clusters (after intrusion)**

| Day | Accepted Users | Total Requests | Percent |
|---|---|---|---|
| 1 | 5 | 10 | 50% |
| 2 | 6 | 10 | 60% |
| 3 | 9 | 20 | 45% |
| 4 | 9 | 20 | 45% |
| 5 | 5 | 10 | 50% |
| 6 | 11 | 20 | 55% |
| 7 | 3 | 9 | 33.33% |
| 8 | 4 | 9 | 44.44% |
| 9 | 2 | 9 | 22.22% |
| 10 | 3 | 9 | 33.33% |
| Total | 57 | 126 | 45.24% |

**(a)** SF1 socialbot

| Day | Accepted Users | Total Requests | Percent |
|---|---|---|---|
| 1 | 8 | 9 | 88.89% |
| 2 | 5 | 9 | 55.56% |
| 3 | 9 | 9 | 100% |
| 4 | 6 | 10 | 60% |
| 5 | 6 | 10 | 60% |
| 6 | 3 | 10 | 30% |
| 7 | 8 | 20 | 40% |
| 8 | 8 | 10 | 80% |
| 9 | 4 | 10 | 40% |
| 10 | 3 | 10 | 30% |
| Total | 60 | 107 | 56.07% |

**(b)** SF2 socialbot

| Day | Accepted Users | Total Requests | Percent |
|---|---|---|---|
| 1 | 5 | 10 | 50% |
| 2 | 1 | 5 | 20% |
| 3 | 7 | 10 | 70% |
| 4 | 4 | 8 | 50% |
| 5 | 4 | 8 | 50% |
| 6 | 2 | 8 | 25% |
| 7 | 2 | 7 | 28.57% |
| 8 | 0 | 0 | 0% |
| 9 | 0 | 0 | 0% |
| 10 | 0 | 0 | 0% |
| Total | 25 | 56 | 44.64% |

**(c)** SF3 socialbot

Table 3: Active socialbots' users accumulation

| Organizations | Before Intrusion | | | | | After Intrusion | | | | | | |
|---|---|---|---|---|---|---|---|---|---|---|---|---|
| | Nodes | Edges | Clusters | | | Nodes | | Edges | | Clusters | | |
| | | | # | Avg. | Max | # | Found (%) | # | Found (%) | # | Avg. | Max |
| $O_{F1}$ | 309 | 1,859 | 35 | 8.6 | 92 | 330 | 6.79 | 2,199 | 18.29 | 34 | 9.5 | 159 |
| $O_{F2}$ | 413 | 3,536 | 19 | 21.421 | 130 | 469 | 13.56 | 3,831 | 8.34 | 29 | 15.828 | 257 |
| $O_{F3}$ | 1,484 | 19,484 | 141 | 9.851 | 268 | 1,675 | 12.87 | 23,823 | 22.27 | 163 | 9.632 | 296 |

Table 4: Intrusion summary

## 4.2.    Reaching Specific Users[16]

In this section we present the results of *Algorithm 2* based on the utilization of our suggested algorithms. The following results are categorized by the OSNs used as well as by the active socialbots. Please notice that accepted users are defined as mutual friends of targeted users who accepted a socialbot's friend request while rejected users

---

[16] The results shown in this section are also published in the paper" Homing Socialbots :Intrusion on a specific organization's employee using Socialbots ",Section V ,and in" Guided Socialbots: Infiltrating the Social Networks of Specific Organizations 'Employees ",Section 4  .

are defined as mutual friends of the targeted users who rejected a socialbot's friend requests.

### 4.2.1. Facebook Results

We used four active socialbots on Facebook in order to infiltrate targeted users in four different organizations. Our socialbots sent friend requests to targeted users' mutual friends in order to gain as many mutual friends as possible to facilitate acceptance by the targeted users. We present the results we received from the $S_{F4}$, $S_{F5}$, and $S_{F6}$ socialbots that attempted to infiltrate specific employees in the targeted organizations, $O_{F4}$, $O_{F5}$, and $O_{F6}$, respectively (see Table 5). The FIS disabled the $S_{F7}$ socialbot, which was intended to infiltrate organization $O_{F7}$.

| Facebook Socialbots | Mutual Friends | Targeted Users |
|---|---|---|
| | Accepted/Total | Accepted/Total |
| $S_{F4}$ | 46/124 | 5/10 |
| $S_{F5}$ | 38/114 | 7/10 |
| $S_{F6}$ | 87/219 | 4/10 |

Table 5: Facebook socialbots' total results

### 4.2.1.1. $S_{F4}$ Socialbot

By the end of the reaching process, $S_{F1}$ sent 124 friend requests to 124 users in the $O_{F4}$ organization, including the ten targeted users. Among them, 46 users accepted, and 78 users rejected $S_{F4}$'s requests (see Figure 9a and 10a).

First, we randomly chose ten users who stated on their Facebook profiles that they work or had worked in the $O_{F4}$ organization. We then collected the friends of the ten targeted users who also worked in the $O_{F4}$ organization and sent them friend requests. Next, socialbot $S_{F4}$ sent friend requests to the ten targeted users ($TU_1$- $TU_{10}$). In total, socialbot $S_{F4}$ sent 124 friend requests and was successful in connecting with 46 different users (see Table 5).

With regard to the targeted users, $S_{F4}$ was able to become friends with five targeted users ($TU_1$, $TU_5$, $TU_6$, $TU_8$, and $TU_{10}$), making for a success rate of 50% (see Table 6, and Figure 11a).

Moreover, $S_{F4}$ was able to become friends with 37.09% of all users who received its friend requests (see Table 6).

| Organization | Targeted Users | Accepted/All Friends | Acceptance Percentage | Accepted? |
|---|---|---|---|---|
| $O_{F4}$ | $TU_1$ | 5/13 | 38.46% | Yes |
|  | $TU_2$ | 4/13 | 30.76% | No |
|  | $TU_3$ | 5/16 | 31.25% | No |
|  | $TU_4$ | 6/16 | 37.50% | No |
|  | $TU_5$ | 6/17 | 35.29% | Yes |
|  | $TU_6$ | 21/42 | 50% | Yes |
|  | $TU_7$ | 7/21 | 33.33% | No |
|  | $TU_8$ | 4/14 | 28.57% | Yes |
|  | $TU_9$ | 7/13 | 53.84% | No |
|  | $TU_{10}$ | 13/32 | 40.62% | Yes |
|  | Total | 46/124 | 37.09% | 50% |

Table 6: OF4 targeted users summary results

### 4.2.1.2.  $S_{F5}$ Socialbot

$S_{F5}$ sent 114 friend requests to 114 users in the $O_{F2}$ organization, including the ten targeted users. Among them, 38 users accepted, and 76 users rejected $S_{F5}$'s requests (see Figure 9b, and 10b).

Regarding the targeted users, socialbot $S_{F5}$ was able to become friends with seven targeted users ($TU_1$, $TU_2$, $TU_3$, $TU_5$, $TU_7$, $TU_9$, and $TU_{10}$), with a success rate of 70% (see Table 7 and Figure 11b).

Moreover, $S_{F5}$ was able to become a friend of 33.33% of all the users who received friend requests (see Table 7).

| Organization | Targeted Users | Accepted/All Friends | Acceptance Percentage | Accepted? |
|---|---|---|---|---|
| $O_{F5}$ | $TU_1$ | 5/12 | 41.16% | Yes |
|  | $TU_2$ | 6/11 | 54.54% | Yes |
|  | $TU_3$ | 7/17 | 41.17% | Yes |
|  | $TU_4$ | 5/16 | 31.25% | No |
|  | $TU_5$ | 11/25 | 44% | Yes |
|  | $TU_6$ | 6/12 | 50% | No |
|  | $TU_7$ | 8/19 | 42.10% | Yes |
|  | $TU_8$ | 6/22 | 27.27% | No |
|  | $TU_9$ | 8/21 | 38.10% | Yes |
|  | $TU_{10}$ | 5/15 | 33.33% | Yes |
|  | Total | 38/114 | 33.33% | 70% |

Table 7: OF5 targeted users summary results

### 4.2.1.3. $S_{F6}$ Socialbot

First, in order to seem like a real user $S_{F6}$ sent 58 friend requests to random users with more than a thousand friends. Among them, 33 users accepted its friend requests, and 19 users asked $S_{F6}$ to be their friend. This means that $S_{F6}$ reached the threshold of fifty users in the first three days (see Figure 8).

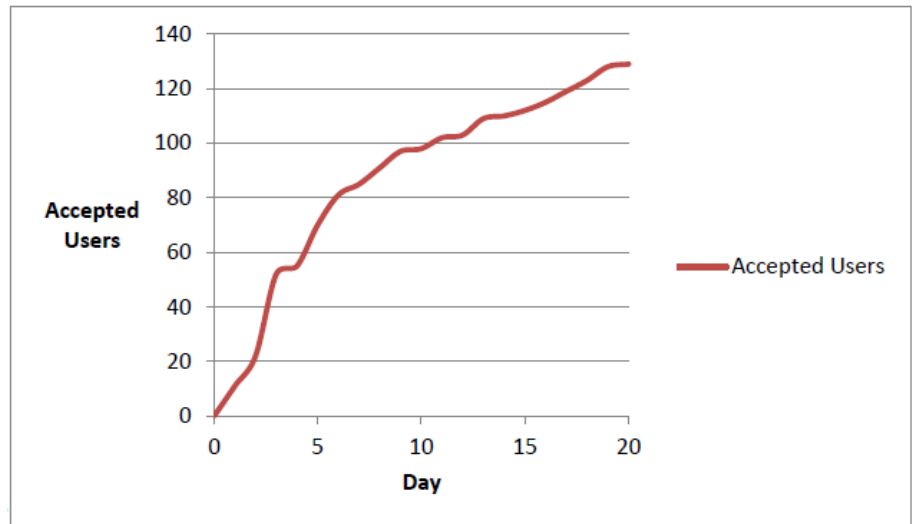

**Figure 8: SF6's random user accumulation**

Overall, $S_{F6}$ accumulated 129 random users as friends and achieved an acceptance rate of 56.9% (see Table 8).

By the end of the reaching process, $S_{F6}$ sent 219 friend requests to 219 users in the $O_{F6}$ organization, including the ten targeted users. Among them, 88 users accepted, and 131 users rejected $S_{F6}$'s requests (see Figure 9c, and 10c).

Regarding the targeted users, $S_{F6}$ was able to become friends with 4 targeted users (TU$_2$, TU$_3$, TU$_8$, and TU$_{10}$) with a success rate of 40% (see Table 9, and Figure 11c). Moreover, $S_{F6}$ was able to become friends with 40.18% of all the users who received friend requests in the infiltration process (see Table 5).

| Day | Accepted Users | | Total | Total Sent | Percent |
|---|---|---|---|---|---|
| | Users Who Got Request | Users Who Sent Request | | | |
| 1 | 11 | 0 | 11 | 17 | 65% |
| 2 | 9 | 2 | 11 | 20 | 45% |
| 3 | 13 | 17 | 30 | 21 | 62% |
| 4 | 0 | 3 | 3 | 0 | |
| 5 | 0 | 15 | 15 | 0 | |
| 6 | 0 | 11 | 11 | 0 | |
| 7 | 0 | 4 | 4 | 0 | |
| 8 | 0 | 6 | 6 | 0 | |
| 9 | 0 | 6 | 6 | 0 | |
| 10 | 0 | 1 | 1 | 0 | |
| 11 | 0 | 4 | 4 | 0 | |
| 12 | 0 | 1 | 1 | 0 | |
| 13 | 0 | 6 | 6 | 0 | |
| 14 | 0 | 1 | 1 | 0 | |
| 15 | 0 | 2 | 2 | 0 | |
| 16 | 0 | 3 | 3 | 0 | |
| 17 | 0 | 4 | 4 | 0 | |
| 18 | 0 | 4 | 4 | 0 | |
| 19 | 0 | 5 | 5 | 0 | |
| 20 | 0 | 1 | 1 | 0 | |
| Total | 33 | 96 | 129 | 58 | 56.90% |

Table 8: SF6's random users accumulation summary results

| Organization | Targeted Users | Accepted/ All Friends | Acceptance Percentage | Accepted? |
|---|---|---|---|---|
| $O_{F6}$ | $TU_1$ | 11/24 | 45.83% | No |
| | $TU_2$ | 17/34 | 50% | Yes |
| | $TU_3$ | 6/22 | 27.27% | Yes |
| | $TU_4$ | 8/18 | 44.44% | No |
| | $TU_5$ | 16/45 | 35.55% | No |
| | $TU_6$ | 19/44 | 43.18% | No |
| | $TU_7$ | 18/42 | 42.85% | No |
| | $TU_8$ | 11/24 | 45.83% | Yes |
| | $TU_9$ | 4/14 | 28.57% | No |
| | $TU_{10}$ | 19/36 | 52.77% | Yes |
| | Total | 87/219 | 39.72% | 40% |

Table 9: OF6 targeted users summary results

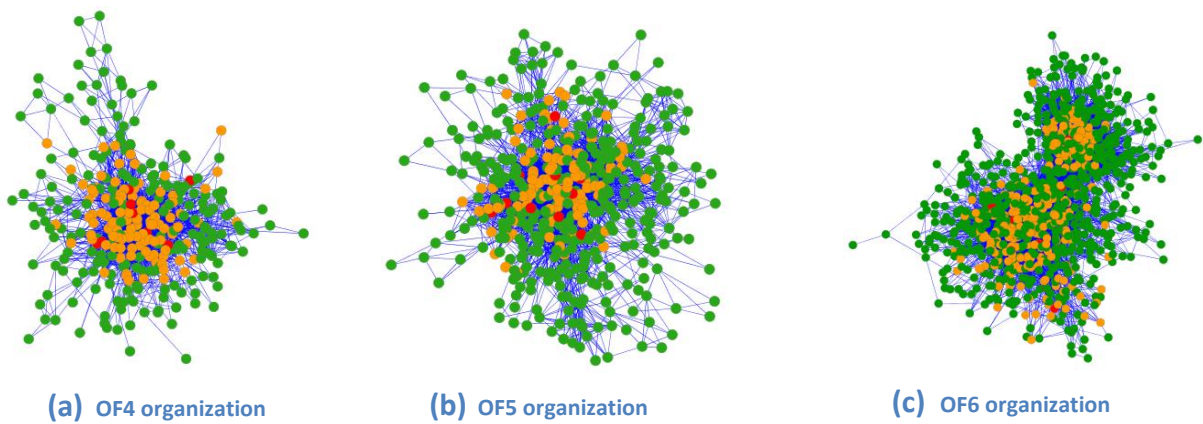

(a) OF4 organization          (b) OF5 organization          (c) OF6 organization

Figure 9: Organizational social network - Red nodes represent targeted users and orange nodes represent users who received friend requests.

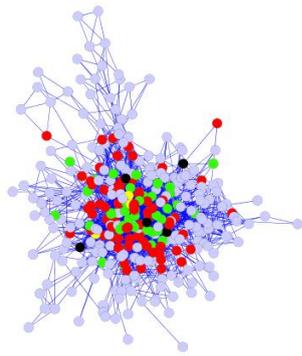 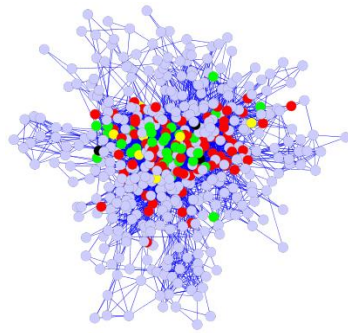 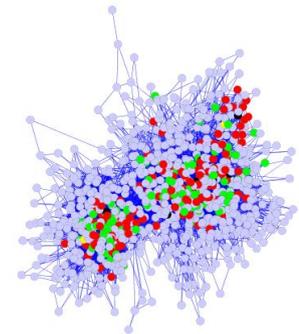

(a) OF4 organization          (b) OF5 organization          (c) OF6 organization

Figure 10: Organizational social network – Green nodes represent accepted users who accepted a friend request from a socialbot, red nodes represent rejected users who did not accept a friend request from a socialbot, yellow nodes represent targeted users who accepted, and black nodes represent targeted users who rejected.

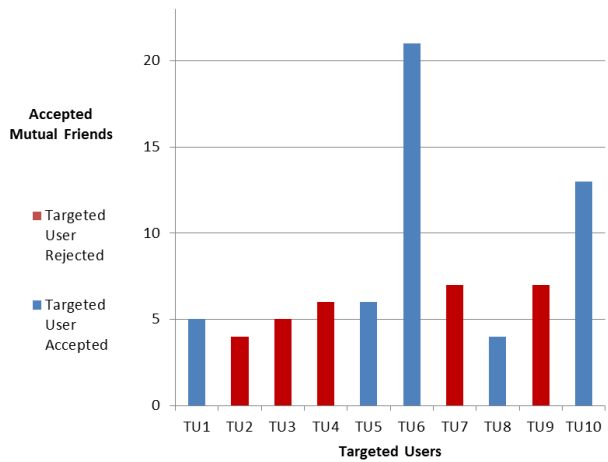

(a) OF4 organization

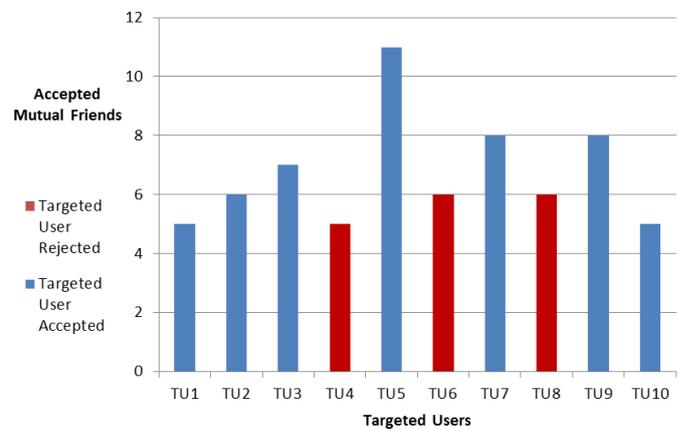

(b) OF5 organization

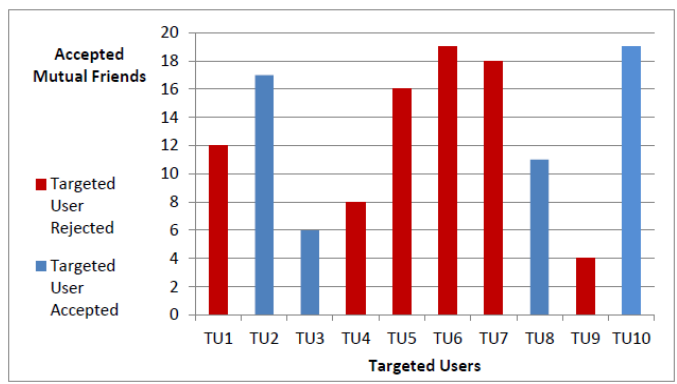

(c) OF6 organization

Figure 11: Organizations' targeted users - The numbers represent how many mutual friends the given socialbot had before it sent a friend request to the targeted users. A blue column represents a targeted user who accepted the socialbot whereas a red column represents a targeted user who rejected the friend request.

### 4.2.2. Xing Results

We used five socialbots on Xing for specific user infiltration of five different organizations. To achieve these goals, our socialbots sent friend requests to targeted users' mutual friends in order to gain as many mutual friends as possible to help gain acceptance by the targeted users. In total, our Xing active socialbots sent 850 friend requests to 850 Xing users. Among them, 439 accepted our socialbots' friend requests (51.64% acceptance rate). We present the results obtained from the Xing socialbots, $S_{x1}$, $S_{x2}$, and $S_{x3}$, which attempted to infiltrate specific employees in the targeted organizations, $O_{x1}$, $O_{x2}$, and $O_{x3}$, respectively (see Table 10). Unfortunately, $S_{x3}$ was blocked prior to completion of the infiltration process so we did not present statistics pertaining to it in Table 10. Moreover, the other two active socialbots, $S_{x4}$ and $S_{x5}$, were blocked by Xing so we present only the results obtained from the accumulation of the random users process (see Section 4.2.2.4, and 4.2.2.5).

| Xing Socialbots | Mutual Friends | Targeted Users |
|---|---|---|
| | Accepted/Total | Accepted/Total |
| $S_{X1}$ | 10/71 | 2/10 |
| $S_{X2}$ | 123/241 | 6/10 |
| $S_{X3}$ | 22/85 | Incomplete |
| $S_{X4}$ | Incomplete | Incomplete |
| $S_{X5}$ | Incomplete | Incomplete |

**Table 10: Xing socialbots summary results**

### 4.2.2.1. $S_{X1}$ Socialbot

First, $S_{x1}$ sent 101 friend requests to random users with more than 400 friends. Among them, 68 users accepted $S_{x1}$'s friend requests, resulting in an acceptance rate of 67.33% in six days (see Table 11, and Figure 12).

| Socialbot | Day | Accepted Users | Total Requests | Percent |
|---|---|---|---|---|
| | 1 | 13 | 20 | 65% |
| | 2 | 11 | 20 | 55% |
| $S_{X1}$ | 3 | 13 | 20 | 65% |
| | 4 | 13 | 20 | 65% |
| | 5 | 17 | 20 | 85% |
| | 6 | 1 | 1 | 100% |
| | Total | 68 | 101 | 67.33% |

**Table 11: SX1 random users summary results**

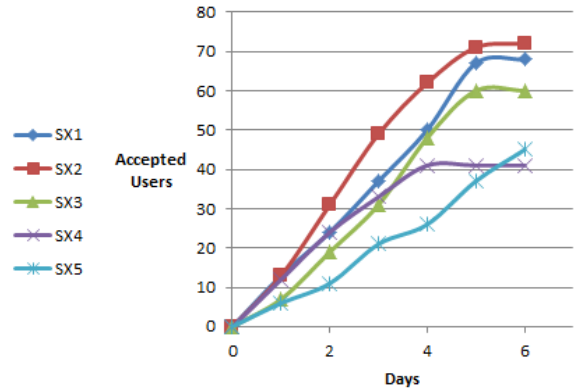

By the end of the infiltration process, $S_{x1}$ sent 71 friend requests to 71 users in the $O_{x1}$ organization, including the ten targeted users. Among them, only ten users accepted (14.08%), and 61 users rejected $S_{x1}$'s requests (see Figures 13a, and 14a).

With regard to targeted users, $S_{x1}$ was able to become friends with two targeted users ($TU_2$, and $TU_5$), with a success rate of 20% (see Table 12, and Figure 15a).

Moreover, $S_{x1}$ was able to become a friend of 14.08% of all users who received friend requests (see Table 12).

| Organization | Targeted Users | Accepted/ All Friends | Acceptance Percentage | Accepted? |
|---|---|---|---|---|
| $O_{x1}$ | $TU_1$ | 1/8 | 12.50% | No |
| | $TU_2$ | 1/11 | 9.09% | Yes |
| | $TU_3$ | 3/14 | 21.42% | No |
| | $TU_4$ | 0/11 | 0.00% | No |
| | $TU_5$ | 2/9 | 22.22% | Yes |
| | $TU_6$ | 1/11 | 9.09% | No |
| | $TU_7$ | 1/8 | 12.50% | No |
| | $TU_8$ | 2/15 | 13.33% | No |
| | $TU_9$ | 4/24 | 16.67% | No |
| | $TU_{10}$ | 0/9 | 0.00% | No |
| | Total | 10/71 | 14.08% | 20% |

**Table 12: OX1 targeted users summary results**

### 4.2.2.2.   $S_{X2}$ Socialbot

$S_{x2}$ sent 87 friend requests to random users with more than 400 friends. Among them, 71 users accepted $S_{x2}$'s friend requests, generating an acceptance rate of 81.61% in five days (see Table 13).

| Socialbot | Day | Accepted Users | | | Total Requests Sent | Percent |
|---|---|---|---|---|---|---|
| | | Users Who Got Request | Users Who Sent Request | Total | | |
| $S_{X2}$ | 1 | 13 | 0 | 13 | 20 | 65% |
| | 2 | 18 | 0 | 18 | 20 | 90% |
| | 3 | 18 | 0 | 18 | 19 | 95% |
| | 4 | 13 | 0 | 13 | 19 | 68% |
| | 5 | 9 | 0 | 9 | 9 | 100% |
| | 6 | 0 | 1 | 1 | 0 | |
| | Total | 71 | 1 | 72 | 87 | 81.61% |

Table 13: SX2 random users summary results

By the end of the infiltration process, $S_{x2}$ sent 241 friend requests to 241 users in the $O_{x2}$ organization (including the ten targeted users). Among them, 123 users accepted, and 118 users rejected $S_{x2}$'s requests (see Figure 13b, and 14b).

With regard to targeted users, $S_{x2}$ was able to become a friend of six targeted users ($TU_3$, $TU_4$, $TU_6$, $TU_7$, $TU_8$, and $TU_{10}$) with a success rate of 60% (see Table 14, and Figure 15b). Moreover, $S_{x2}$ was able to become a friend of 51.04% of all users who received friend requests (see Table 14).

| Organization | Targeted Users | Accepted/ All Friends | Acceptance Percentage | Accepted? |
|---|---|---|---|---|
| $O_{X2}$ | $TU_1$ | 9/24 | 37.5% | No |
| | $TU_2$ | 20/41 | 48.78% | No |
| | $TU_3$ | 6/12 | 50% | Yes |
| | $TU_4$ | 10/29 | 34.48% | Yes |
| | $TU_5$ | 8/19 | 42.10% | No |
| | $TU_6$ | 18/27 | 66.67% | Yes |
| | $TU_7$ | 18/26 | 69.23% | Yes |
| | $TU_8$ | 20/41 | 48.78% | Yes |
| | $TU_9$ | 15/30 | 50% | No |
| | $TU_{10}$ | 19/34 | 55.88% | Yes |
| | Total | 123/241 | 51.04% | 60% |

Table 14: OX2 targeted users summary results

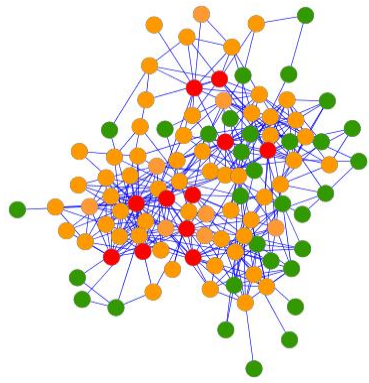

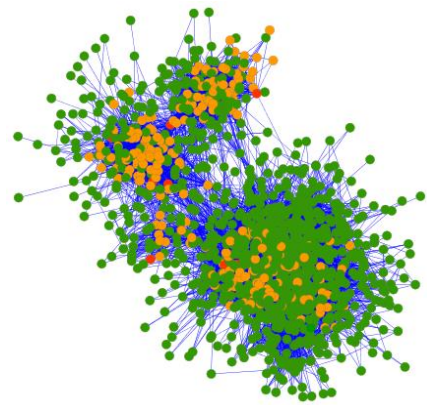

(a) OX1 organization                          (b) OX2 organization

Figure 13: Organizational social network - Red nodes represent targeted users and orange nodes represent users who received friend requests.

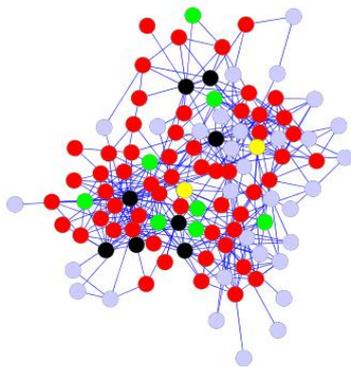

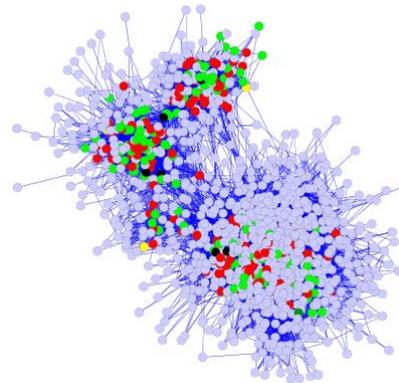

(a) OX1 organization                          (b) OX2 organization

Figure 14: Organizational social network – Green nodes represent accepted users who accepted a friend request from a socialbot, red nodes represent rejected users who did not accept friend request, yellow nodes represent targeted users who accepted, and black nodes represent targeted users who rejected.

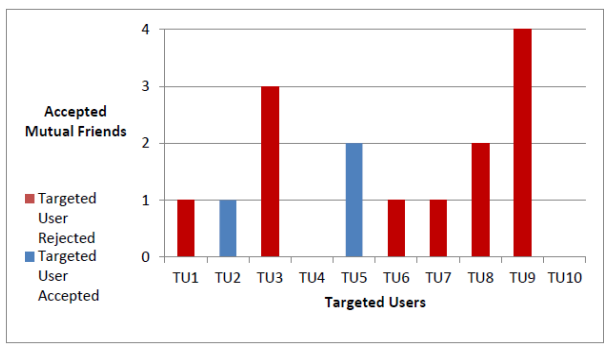

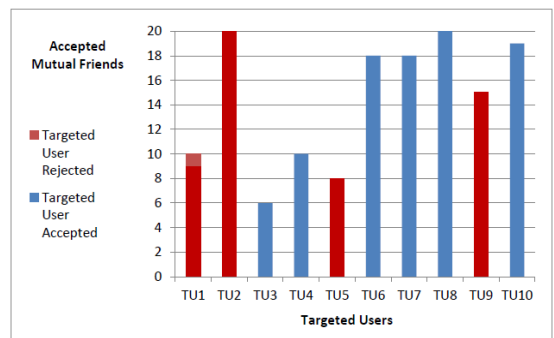

(a) OX1 organization                          (b) OX2 organization

Figure 15: Organizations' targeted users - The numbers represent how many mutual friends the given socialbot had before it sent a friend request to the targeted users. A blue column represents a targeted user who accepted the socialbot whereas a red column represents a targeted user who rejected the friend request.

### 4.2.2.3.  $S_{X3}$ Socialbot

In order to seem as if it were an actual user, $S_{x3}$ sent 90 friend requests to random users with more than 400 friends. Among them, 59 users accepted $S_{x3}$'s friend requests, creating an acceptance rate of 65.56% in five days (see Table 15).

| Socialbot | Day | Accepted Users | | | Total Requests Sent | Percent |
|---|---|---|---|---|---|---|
| | | Users Who Got Request | Users Who Sent Request | Total | | |
| $S_{X3}$ | 1 | 7 | 0 | 7 | 15 | 47% |
| | 2 | 12 | 0 | 12 | 20 | 60% |
| | 3 | 12 | 0 | 12 | 20 | 60% |
| | 4 | 17 | 0 | 17 | 20 | 85% |
| | 5 | 11 | 1 | 12 | 15 | 73% |
| | Total | 59 | 1 | 60 | 90 | 65.56% |

Table 15: SX3 random users summary results

By the end of the infiltration process, $S_{x3}$ had sent 85 friend requests to 85 users in the $O_{x3}$ organization before it was blocked by Xing. Among the users who received friend requests, 22 users accepted and 63 users rejected them—making for an acceptance rate of 25.88% in five days (see Table 16).

| Organization | Targeted Users | Accepted/ All Friends | Acceptance Percentage | Accepted? |
|---|---|---|---|---|
| $O_{X3}$ | $TU_1$ | 5/11 | 45.5% | No |
| | $TU_2$ | 3/10 | 30% | |
| | $TU_3$ | 7/29 | 24.13% | |
| | $TU_4$ | 6/19 | 31.57% | |
| | $TU_5$ | 4/15 | 26.67% | |
| | $TU_6$ | 2/11 | 18.18% | |
| | $TU_7$ | 2/12 | 16.67% | |
| | $TU_8$ | 2/21 | 9.52% | |
| | $TU_9$ | 1/12 | 8.33% | |
| | $TU_{10}$ | 3/9 | 33.33% | |
| | Total | 22/85 | 25.88% | |

Table 16: OX3 targeted users summary results

### 4.2.2.4.  $S_{X4}$ Socialbot

$S_{x4}$ did not manage to reach the threshold of fifty friends when sending friend requests to random users with more than 400 friends (see Figure 12), as in the middle of the process it was blocked by Xing. $S_{x4}$ did manage to send 75 friend requests to random users with more than 400 friends before it was blocked. Among them, 41 users accepted $S_{x4}$'s friend requests—producing an acceptance rate of 54.67% in four days (see Table 17).

| Socialbot | Day | Accepted Users | Total Requests | Percent |
|-----------|-----|----------------|----------------|---------|
| $S_{X4}$ | 1 | 12 | 20 | 60% |
| | 2 | 12 | 20 | 60% |
| | 3 | 9 | 20 | 45% |
| | 4 | 8 | 15 | 53% |
| | Total | 41 | 75 | 54.67% |

Table 17: SX4 random users summary results

### 4.2.2.5.   $S_{X5}$ Socialbot

$S_{x5}$ was not able to reach the threshold of 50 friends when sending requests to users with more than 400 friends (see Figure 12). As was the case with $S_{x5}$, in the middle of the infiltration process, $S_{x5}$ was blocked by Xing. $S_{x5}$ managed to send 100 friend requests to random users before it was blocked, and 45 users accepted $S_{x5}$'s friend requests. This made for an acceptance rate of 45% in six days (see Table 18).

| Socialbot | Day | Accepted Users | Total Requests | Percent |
|-----------|-----|----------------|----------------|---------|
| $S_{X5}$ | 1 | 6 | 15 | 40% |
| | 2 | 5 | 10 | 50% |
| | 3 | 10 | 15 | 67% |
| | 4 | 5 | 20 | 25% |
| | 5 | 11 | 20 | 55% |
| | 6 | 8 | 20 | 40% |
| | Total | 45 | 100 | 45% |

Table 18: SX5 random users summary results

# 5. Ethical Considerations

Today, most OSNs do not allow free access to personal information due to the privacy concerns of network users and the OSN's terms of use [77]. As a result, much of the research using OSNs involves various techniques of collecting sensitive data by circumventing OSN limitations. Elovici et al. [77] performed a comprehensive review of research involving OSNs. They described two kinds of OSN research: "Whitehat" research that is defined as legitimate academic and industrial investigation, and "Blackhat/greyhat" research that is defined as studying and exploiting vulnerabilities of OSNs in order to extract sensitive information, actively connect to users, create fake identities, and even perform malicious activities. The actions that researchers have to do in order to get this kind of data are controversial and raise many ethical questions.

In order to perform accurate blackhat/greyhat research, a researcher must base his or her study on actual OSN data, such as real connections between users, correct textual content, authentic files, etc. Researchers must monitor many real-life OSN users in order to study the diffusion of data in OSNs. An effective and widely employed technique to obtain data from OSNs and their users is based on establishing connections with users, typically by creating a large number of fake OSN user accounts, which are then used to connect other users.

Moreover, Elovici et al. claimed that the main goal of academic blackhat/greyhat research is done with a precise purpose: to study vulnerabilities in order to create improved defenses for OSNs and their users.

Given the very nature of OSNs, we should ask the following question: Is it ethical to perform research such as ours? We believe that the answer to this question is affirmative for several reasons.

First, during the period that we conducted this study, Ben-Gurion University of the Negev, did not question our work or require approval by the research ethics committee in order to conduct the study. Furthermore, we made a great effort in order to increase the standard of ethics in the domain of OSNs [10], [77], [78].

Second, in the initial crawling process of our study on Facebook and Xing, we collected only publicly available data that was accessible to every registered user.

Third, we avoided using profile images of real users when creating the identities of our socialbots. Instead, we selected profile images that either did not include users' faces or that presented the faces in such a way that it was impossible to identify the person, or we Photoshopped images of fictional entities.

Fourth, although this study included actual OSN users, and the results may inadvertently provide knowledge of concern to OSN users and operators, ignoring the problem does not provide a solution. We can rid ourselves of responsibility for these challenging situations and choose not to perform such research, however, the problem will continue to exist and in fact, increase. Performing this type of research aids the development of new forms of protection provided by OSNs.

Lastly, given the enormous number of OSN users and the extensive opportunities to exploit the personal information of each and every one of them, it is crucial to study the dangers and privacy issues that exist for users of OSNs. As was previously mentioned, Facebook itself has estimated that 8.7% of its accounts are defined as fake [45], and certainly some percentage of these are malicious. OSN users need to be aware of the online dangers that exist and modify their actions accordingly. Online security represents an acute problem that must be studied by legitimate researchers in order to be effectively addressed by industry and the academia.

# 6. Discussion

Using the methods described in Section 3.2, our socialbots were able to reach specific employees in three different targeted organizations within the Facebook OSN and two targeted organizations within the Xing OSN. However, we should extend our discussion beyond these successful infiltrations.

First, there are points of comparison between the socialbots that completed the reaching process successfully and the ones that were exposed and eventually blocked by the OSNs. In total, we operated four socialbots on Facebook. Three of four socialbots completed this process. $S_{F4}$, $S_{F5}$, and $S_{F6}$ achieved success rates of 50%, 70%, and 40% respectively (see Table 5). Socialbot $S_{F7}$ conversely, was blocked by Facebook operators in the middle of the process. We believe that the failure of $S_{F7}$ lies in the location feature: namely, the fact that there was a great geographic distance between the targeted organization $O_{F7}$ and $S_{F7}$ socialbot's current location attribute. The air travel distance between the location of most of the $O_{F7}$ organization employees and $S_{F7}$'s current location was more than 4,000 kilometers, spanning several countries. It is important to note that the difference between the targeted organization's employees and the socialbot's current location existed only in the case of $O_{F7}$; in all other cases, the location of the socialbot was similar to the location of the targeted organization. We can assume from this incident that there were likely cultural differences—which logically correspond to geographic distance—between OSN users, and such differences could raise suspicions when contemplating whether to accept friend requests. We hope to verify this assumption in a future study. Therefore, we decided to tailor the socialbot's identity to its target. Moreover, this incident reinforces the conclusion of Liben-Nowell et al. [79] regarding their finding of a strong correlation between friendship and geographic location on LiveJournal.[18]

With regard to our infiltration of Xing, we operated five socialbots. Among them, two of the five completed the reaching process. $S_{X1}$, and $S_{X2}$ succeeded in reaching specific employees; however, $S_{X3}$, $S_{X4}$, and $S_{X5}$ failed in their respective missions (see Sections 4.2.2.3, 4.2.2.4, and 4.2.2.5). We believe that the failure in these three cases resulted from the socialbot's organizational affiliation. Our successful socialbots, $S_{X1}$ and $S_{X2}$, were defined as users who were not connected directly to the targeted organizations: $S_{X1}$ was defined as a freelancer coach, whereas $S_{X2}$ was defined as a recruiter in a technology-oriented organization. They were able to infiltrate specific employees in the targeted organization with the success rate of 20% and 60% (see Table 10). In contrast to these two socialbots, we defined $S_{X3}$ as an employee within the $O_{X3}$ organization. $S_{X3}$ succeeded in passing the first stage, the accumulation of fifty random users (see Table 15), but once $S_{X3}$ began initiating friend requests to employees of $O_{X3}$, suspicions arose. Several suspicious users accessed their organization's human resource software to verify $S_{X3}$'s false identity. Some of them even notified the socialbot that they knew for sure that the identity was not an employee in their organization. A few days later, the $S_{X3}$ socialbot was blocked by Xing. These actions taken by the employees using organizational software tools for recognition, prior to accepting a friend request demonstrates how a well-formed policy and clear instructions to employees can benefit the security of employees and their organizations on OSNs.

---

[18] http://www.livejournal.com

The socialbots $S_{X4}$ and $S_{X5}$ failed in the middle phase of the accumulation of fifty random users (see Tables 17, and 18). We defined these two socialbots as employees in two random organizations. It is important to mention that these two random organizations were not the targeted organizations that we wanted to infiltrate, but separate organizations we had chosen for the identity of the socialbots. While trying to evade detection by Xing operators based on the community's structure of OSN users (see Section 3.4.3), our socialbot's friend requests became too embedded within the random organizations: Among the 75 friend requests that were sent by $S_{X4}$, 28 users were employees in the randomly chosen organization (37%). Similarly, among 100 friend requests that were sent by $S_{X5}$, 58 users were employees in the randomly chosen organization (58%). $S_{X4}$ and $S_{X5}$ were blocked by Xing operators when employees from these organizations verified the identities of these two socialbots.

Another issue we wish to discuss is the low infiltration of Xing socialbot $S_{X1}$, which was able to infiltrate only 20% of the targeted users (see Tables 10 and 12). We believe that this poor performance was related to the fact of the 61 users who did not respond to $S_{X1}$'s friend requests, there were 25 users who had 0% activity. This means that 41% of the users who did not respond to $S_{X1}$'s friend requests were inactive, i.e, users who did not actually receive the friend requests. It is important to understand that we made a point not to avoid these zero-activity users in our study; we sent friend requests to users who had 0% activity regardless, in order to fairly evaluate their role.

Lastly, we want to consider the recommended threshold of the number of mutual friends that would influence a specific user to accept our socialbot's friend request. Our previous study [13] found that a socialbot would typically be accepted as a friend of specific employees when it had gained seven or more mutual friends of the targeted user. This number of mutual friends corresponds to the results illustrated in Figures 11a and 11b. However, when expanding the study to include one more organization on Facebook (see Figure 11c) and two organizations on Xing (see Figures 15a and 15b), we can suggest that the threshold has increased to 17 or more mutual friends, and the probability that a targeted user will accept our socialbot's request is 70%. There were 7 targeted users with 17 or more mutual friends who accepted our socialbots' friend requests: one targeted user from $O_{F4}$, two targeted users from $O_{F6}$, and four targeted users from $O_{X2}$. There were three targeted users with 17 or more mutual friends who rejected our socialbots' friend requests: two targeted users from $O_{F6}$, and one targeted user from $O_{X2}$. These suggestions are further reinforced by the conclusions of Boshmaf et al. [1] that the more a user's mutual friends accept the socialbot's requests, the more likely the targeted user is to accept the socialbot's friend request as well.

# 7. Conclusion

In this thesis, we demonstrated two attacks by socialbots. First, we demonstrated how adversaries may use socialbots for mining data regarding targeted organizations. Second, we demonstrated how adversaries may reach specific employees in targeted organizations within OSNs. This further emphasizes the persistent privacy issues surrounding OSNs. Based on our results, we can draw the following conclusions and recommendations.

First, with regard to the data mining of targeted organizations our proposed method was perceived as a great success. By accepting the socialbots' friend requests, we were able to expose more information, such as hidden employees, connections between employees, and groups, in comparison with the public information we gathered with the passive socialbot P. With regard to hidden employees, $S_{F1}$ found 330 employees compared with P which found 309 employees, $S_{F2}$ found 469 employees compared with P which found 413 employees, and $S_{F3}$ found 1,675 employees compared with P which found 1,484 employees. Regarding hidden connections, $S_{F1}$ found 2,199 links compared with P which found 1,859 links, $S_{F2}$ found 3,831 links compared with P which found 3,536 links, and $S_{F3}$ found 23,823 links compared with P which found 19,484 links (see Table 4).

Moreover, we were able to expose a different number of clusters with our socialbots: 34 clusters were found by $S_{F1}$ - compared to 35 clusters found by P, 29 clusters were found by $S_{F2}$ compared with 19 clusters found by P, and 163 clusters were found by $S_{F3}$ compared with 141 clusters found by P (see Table 4). However by using other measures, like the closeness centrality measure, we received the same number of clusters with both the passive and the active socialbots. Finally, we can conclude that using our active socialbots, we succeeded in discovering up to 13.55% more employees and up to 23.304% more informal organizational links when compared to the organizations' social network collected by the passive socialbots (see Table 4). Our results demonstrate that organizations that are interested in protecting themselves should instruct their employees not to disclose information on social networks and to be careful when accepting friend requests.

Second, we were able to show that OSN users still tend to accept friend requests from complete strangers. Most of our socialbots on both OSNs were able to reach the threshold of fifty random users within 5-6 days (see Figures 5, 8, and 12). OSN users should realize how easy it is for an attacker to create socialbots, and how insignificant it is to attackers when their socialbots are blocked. In the case that a socialbot is blocked, the attacker can quickly create a new, improved fake profile and continue with the infiltration process. OSN users must understand the risks of accepting a friend request from people they do not know and should ignore friend requests received from strangers.

Third, our experimental results from the process of reaching specific employees indicate that there is a link between having mutual friends and the acceptance of friend requests. The step of first being accepted as a friend of a mutual friend of the targeted users in targeted organizations was significant to our socialbots' ability to infiltrate. Without this step, we believe that the acceptance rate would have been much lower. Moreover, we found that if a socialbot had 17 or more mutual friends of the targeted user, the probability that the targeted user would accept the socialbot's friend request was 70% (see Section 6).

Fourth, malicious socialbots operate within OSNs and can function at a high level of sophistication. As we learned from executing our suggested algorithm, most of our socialbots were able to infiltrate specific employees in targeted organizations on both Facebook and Xing, despite the differences between these two OSNs. Please note that we intend to further explore cases of socialbot blocking in a future study. Furthermore, our results of the $S_{F7}$ socialbot may indicate that users tend to trust strangers on the basis of their mutual attributes like current location, mutual friends, etc (see Section 6). We recommend that users not rely upon these mutual features when a stranger sends them a friend request.

Fifth, organizations should understand the risks of organizational information leakage that might occur due to their employees using OSNs. Moreover, we strongly recommend that organizations should take responsibility for raising the level of awareness of employees to this problematic phenomenon and for underscoring the risks posed to employees when they accept unfamiliar users as friends. Organizations should explain to OSN users the risks of data leakage and provide them with tools to verify users who declare themselves to be employees in the organizations. This kind of software could help employees verify whether or not a stranger who sends them a friend request is a real employee. This recommendation also endorses a recommendation by Fire et al. [10] that suggested performing a short security check on a stranger.

This study has several future research directions. One possible direction is more thorough testing of the conclusions we found regarding the cultural differences between users and organizational affiliation of socialbots when we define their identity. Another possible direction is to use the algorithm for reaching specific users to investigate whether our results are consistent over time and to assess whether there are changes in users' awareness and responses to privacy issues. Moreover, we can use the algorithm on other OSNs and observe the differences between them. Furthermore, we could differentiate between female and male profiles when reaching OSN users to investigate any gender differences that exist.

In both the present and the future, individuals and organizations need to be aware that harmful socialbots exist on OSNs. Consequently, users must access social networks wisely and should establish effective security and privacy measures.

# References

[1]  Y. Boshmaf, I. Muslukhov, K. Beznosov and M. Ripeanu, "The Socialbot Network: When Bots Socialize for Fame and Money," in Proceedings of the 27th Annual Computer Security Applications Conference, ACM, 2011, pp. 93-102.

[2]  A. O'cass and T. Fenech, "Web retailing adoption: exploring the nature of internet users Web retailing behaviour," Journal of Retailing and Consumer services, vol. 10, no. 2, p. 81–94, 2003.

[3]  E. F. Gross, "Adolescent Internet use: What we expect, what teens report," Journal of Applied Developmental Psychology, vol. 25, no. 6, pp. 633-649, 2004.

[4]  A. Lenhart, K. Purcell, A. Smith and K. Zickuhr, "Social Media & Mobile Internet Use among Teens and Young Adults. Millennials," Pew Internet & American Life Project, 2010.

[5]  J. A. Diaz, R. A. Griffith, J. J. Ng, S. E. Reinert, P. D. Friedmann and A. W. Moulton, "Patients' Use of the Internet for Medical Information," Journal of general internal medicine, vol. 17, no. 3, pp. 180-185, 2002.

[6]  D. M. Boyd and N. B. Ellison, "Social Network Sites: Definition, History, and Scholarship," Journal of Computer-Mediated Communication, vol. 13, no. 1, pp. 210-230, 2007.

[7]  R. E. Wilson, S. D. Gosling and L. T. Graham, "A Review of Facebook Research in the Social Sciences," Perspectives on Psychological Science, vol. 7, no. 3, pp. 203-220, 2012.

[8]  L. Bilge, T. Strufe, D. Balzarotti and E. Kirda, "All Your Contacts Are Belong to Us: Automated Identity Theft Attacks on Social Networks," in Proceedings of the 18th international conference on World wide web, ACM, 2009, pp. 551-560.

[9]  K. Manning, "The impacts of Online Social Networking and Internet Use on Human Communication and Relationships," 2009.

[10] M. Fire, R. Goldschmidt and Y. Elovici, "Online Social Networks: Threats and Solutions," Communications Surveys & Tutorials, IEEE, vol. 16, no. 4, pp. 2019-2036, 2014.

[11] M. Fire and R. Puzis, "Organization Mining Using Online Social Networks," in Networks and Spatial Economics, Springer, 2012, pp. 1-34.

[12] A. Elyashar, F. Michael, D. Kagan and Y. Elovici, "Organizational intrusion: Organization mining using socialbots," in Social Informatics (SocialInformatics), 2012 International Conference on, IEEE, 2012, pp. 7-12.

[13] A. Elyashar, M. Fire, D. Kagan and Y. Elovici, "Homing Socialbots: Intrusion on a specific organizations employee using Socialbots," in Proceedings of the 2013 IEEE/ACM International Conference on Advances in Social Networks Analysis and Mining, ACM, 2013, pp. 1358-1365.

[14] A. Elyashar, M. Fire, D. Kagan and Y. Elovici, "Guided Socialbots: Infiltrating User's Friends List," in AI Communications, IOS Press, 2014.

[15] Y. Zhen, "A Novel Spam Campaign in Online Social Networks," 2013.

[16] Y. Boshmaf, I. Muslukhov, K. Beznosov and M. Ripeanu, "Design and Analysis of a Social Botnet," Computer Networks, vol. 57, no. 2, pp. 556-578, 2013.

[17] A. Paradise, R. Puzis and A. Shabtai, "Anti-Reconnaissance Tools: Detecting Targeted Socialbots," Internet Computing, IEEE, vol. 18, no. 5, pp. 11-19, 2014.

[18] C. Wagner, S. Mitter, C. Korner and M. Strohmaier, "When social bots attack: Modeling susceptibility of users in online social networks," in Making Sense of Microposts (MSM2012), Citeseer, 2012, p. 2.

[19] C. I. Nombo, When AIDS meets poverty, vol. 5, Wageningen Academic Publishers, AWLAE Series, 2008.

[20] J. Ratkiewicz, M. Conover, M. Meiss, B. Goncalves, S. Patil, A. Flammini and F. Menczer, "Detecting and tracking the spread of astroturf memes in microblog streams," arXiv preprint arXiv:1011.3768, 2010.

[21] "Facebook Newsroom," [Online]. Available: http://newsroom.fb.com/company-info/. [Accessed 29 August 2015].

[22] "Our List Of The World's Largest Social Networks Shows How Video, Messages, And China Are Taking Over The Social Web," 2013. [Online]. Available: http://www.businessinsider.com/the-worlds-largest-social-networks-2013-12.

[23] R. Gross and A. Acquisti, "Information revelation and privacy in online social networks," in Proceedings of the 2005 ACM workshop on Privacy in the electronic society, ACM, 2005, pp. 71-80.

[24] S. B. Barnes, "A privacy paradox: Social networking in the United States," First Monday, vol. 11, no. 9, 2006.

[25] J. Lindamood, R. Heatherly, M. Kantarcioglu and B. Thuraisingham, "Inferring private information using social network data," in Proceedings of the 18th international conference on World wide web, ACM, 2009, pp. 1145-1146.

[26] S. Mahmood, "Online Social Networks: Privacy Threats and Defenses," in Security and Privacy Preserving in Social Networks, Springer, 2013, pp. 47-71.

[27] J. Lindamood, R. Heatherly, M. Kantarcioglu and B. Thuraisingham, "Inferring Private Information Using Social Network Data," in Proceedings of the 18th international conference on World wide web, ACM, 2009, pp. 1145-1146.

[28] S. Ghosh, B. Viswanath, F. Kooti, N. K. Sharma, G. Korlam, F. Benevenuto, N. Ganguly and K. P. Gummadi, "Understanding and Combating Link Farming in the Twitter Social Network," in Proceedings of the 21st international conference on World Wide Web, ACM, 2012, pp. 61-70.

[29] A. Mislove, B. Viswanath, K. P. Gummadi and P. Druschel, "You Are Who You Know: Inferring User Profiles in Online Social Networks," in Proceedings of the third ACM international conference on Web search and data mining, ACM, 2010, pp. 251-260.

[30] J. Baltazar, J. Costoya and R. Flores, "The Real Face of KOOBFACE: The Largest Web 2.0 Botnet Explained," Trend Micro Research, vol. 5, no. 9, p. 10, 2009.

[31] Q. Cao, M. Sirivianos, X. Yang and T. Pregueiro, "Aiding the Detection of Fake Accounts in Large Scale Social Online Services," in Proceedings of the 9th USENIX conference on Networked Systems Design and Implementation, USENIX Association, 2012, p. 15.

[32] G. Stringhini, G. Wang, M. Egele, C. Kruegel, G. Vigna, H. Zheng and B. Y. Zhao, "Follow the Green: Growth and Dynamics in Twitter Follower Markets," in Proceedings of the 2013 conference on Internet measurement conference, ACM, 2013, pp. 163-176.

[33] Z. Chu, S. Gianvecchio, H. Wang and S. Jajodia, "Who is Tweeting on Twitter: Human, Bot, or Cyborg?," in Proceedings of the 26th annual computer security applications conference, ACM, 2010, pp. 21-30.

[34] J. Wolak, D. Finkelhor, K. J. Mitchell and M. L. Ybarra, "Online "predators" and their victims: myths, realities, and implications for prevention and treatment," American Psychologist, vol. 63, no. 2, p. 111, 2008.

[35] M. L. Ybarra and K. J. Mitchell, "How Risky Are Social Networking Sites? A Comparison of Places Online Where Youth Sexual Solicitation and Harassment Occurs," Pediatrics, vol. 121, no. 2, pp. 350-357, 2008.

[36] M. Balduzzi, C. Platzer, T. Holz, E. Kirda, D. Balzarotti and C. Kruegel, "Abusing Social Networks for Automated User Profiling," in Recent Advances in Intrusion Detection, Springer, 2010, pp. 422-441.

[37] J. Lewis and S. Baker, "The Economic Impact of Cybercrime and Cyber Espionage," Center for Strategic and International Studies, Washington, DC, 2013.

[38] J. Kostka, Y. A. Oswald and R. Wattenhofer, "Word of Mouth: Rumor Dissemination in Social Networks," in Structural Information and Communication Complexity, Springer, 2008, pp. 185-196.

[39] M. Nekovee, Y. Moreno, G. Bianconi and M. Marsili, "Theory of rumour spreading in complex social networks," Physica A: Statistical Mechanics and its Applications, vol. 374, no. 1, pp. 457-470, 2007.

[40] K. Peterson and K. A. Siek, "Analysis of Information Disclosure on a Social Networking Site," in Online Communities and Social Computing, Springer, 2009, pp. 256-264.

[41] J. Ratkiewicz, M. Conover, M. Meiss, B. Goncalves, A. Flammini and F. Menczer, "Detecting and Tracking Political Abuse in Social Media," in ICWSM, 2011.

[42] J. Ratkiewicz, M. Conover, M. Meiss, B. Goncalves, S. Patil, A. Flammini and F. Menczer, "Truthy: Mapping the Spread of Astroturf in Microblog Streams," in Proceedings of the 20th international conference companion on World wide web, ACM, 2011, pp. 249-252.

[43] J. Arquilla and D. Ronfeldt, "Networks and netwars: The future of terror, crime, and militancy," Rand Corporation, 2001.

[44] J. Ugander, B. Karrer, L. Backstrom and C. Marlow, "The anatomy of the facebook social graph," arXiv preprint arXiv:1111.4503, 2011.

[45] "UNITED STATES SECURITIES AND EXCHANGE COMMISSION," 2012.

[46] T. Stein, E. Chen and K. Mangla, "Facebook Immune System," in Proceedings of the 4th Workshop on Social Network Systems, ACM, 2011, p. 8.

[47] Y. Liu, K. P. Gummadi, B. Krishnamurthy and A. Mislove, "Analyzing facebook privacy settings: user expectations vs. reality," in Proceedings of the 2011 ACM SIGCOMM conference on Internet measurement conference, ACM, 2011, pp. 61-70.

[48] "Xing Corporate Pages," Xing, [Online]. Available: https://corporate.xing.com/no_cache/english/company/xing-ag/. [Accessed 29 August 2015].

[49] "Twealing Social Networks," [Online]. Available: https://tweakingsocialnetworks.wordpress.com/tag/bookmark-unconfirmed-contacts-in-xing/. [Accessed 29 August 2015].

[50] C. Dwyer, S. R. Hiltz and K. Passerini, "Trust and Privacy Concern Within Social Networking Sites: A Comparison of Facebook and MySpace," AMCIS 2007 Proceedings, p. 339, 2007.

[51] T. Ryan and G. Mauch, "Getting In bed with Robin Sage," in Black Hat Conference, 2010.

[52] D. H. Chau, S. Pandit, S. Wang and C. Faloutsos, "Parallel Crawling for Online Social Networks," in Proceedings of the 16th international conference on World Wide Web, ACM, 2007, pp. 1283-1284.

[53] H. Kwak, C. Lee, H. Park and S. Moon, "What is Twitter, a Social Network or a News Media?," in Proceedings of the 19th international conference on World wide web, ACM, 2010, pp. 591-600.

[54] R. T. Stern, L. Samama, R. Puzis, T. Beja, Z. Bnaya and A. Felner, "TONIC: Target Oriented Network Intelligence Collection for the Social Web," in AAAI, 2013.

[55] N. M. Tichy, M. L. Tushman and C. Fombrun, "Social network analysis for organizations," Academy of management review, vol. 4, no. 4, pp. 507-519, 1979.

[56] V. E. Krebs, "Mapping networks of terrorist cells," Connections, vol. 24, no. 3, pp. 43-52, 2002.

[57] N. Mishra, R. Schreiber, I. Stanton and R. E. Tarjan, "Clustering Social Networks," in Algorithms and Models for the Web-Graph, Springer, 2007, pp. 56--67.

[58] "Sophos Facebook ID probe," 2007. [Online]. Available: http://www.sophos.com/en-us/press-office/press-releases/2007/08/facebook.aspx. [Accessed 29 August 2015].

[59] J. Bonneau, J. Anderson and G. Danezis, "Prying Data out of a Social Network," in Social Network Analysis and Mining, 2009. ASONAM'09. International Conference on Advances in, IEEE, 2009, pp. 249-254.

[60] M. Magdon-Ismail and B. Orecchio, "Guard your connections: infiltration of a trust/reputation based network," in Proceedings of the 4th Annual ACM Web Science Conference, ACM, 2012, pp. 195-204.

[61] S. Mitter, C. Wagner and M. Strohmaier, "A categorization scheme for socialbot attacks in online social networks," Paris, France, ACM, 2013.

[62] S. Mitter, C. Wagner and M. Strohmaier, "Understanding the impact of socialbot attacks in online social networks," arXiv preprint arXiv:1402.6289, 2014.

[63] C. A. Freitas, F. Benevenuto, S. Ghosh and A. Veloso, "Reverse engineering socialbot infiltration strategies in twitter," arXiv preprint arXiv:1405.4927, 2014.

[64] F. Benevenuto, G. Magno, T. Rodrigues and V. Almeida, "Detecting Spammers on Twitter," Collaboration, electronic messaging, anti-abuse and spam conference (CEAS), vol. 6, p. 12, 2010.

[65] M. Fire, G. Katz and Y. Elovici, "Strangers Intrusion Detection - Detecting Spammers and Fake Profiles in Social Networks Based on Topology Anomalies," HUMAN, vol. 1, no. 1, p. 26, 2012.

[66] K. Lee, J. Caverlee and S. Webb, "Uncovering Social Spammers: Social Honeypots + Machine Learning," in Proceedings of the 33rd international ACM SIGIR conference on Research and development in information retrieval, ACM, 2010, pp. 435-442.

[67] G. Stringhini, C. Kruegel and G. Vigna, "Detecting Spammers on Social Networks," in Proceedings of the 26th Annual Computer Security Applications Conference, ACM, 2010, pp. 1-9.

[68] S. Blackmore, "Imitation and the definition of a meme," Journal of Memetics-Evolutionary Models of Information Transmission, vol. 2, no. 11, pp. 159-170, 1998.

[69] G. Wang, T. Konolige, C. Wilson, X. Wang, H. Zheng and B. Y. Zhao, "You Are How You Click: Clickstream Analysis for Sybil Detection," in Usenix Security, 2013, pp. 241-256.

[70] R. Wald, T. M. Khoshgoftaar, A. Napolitano and C. Sumner, "Predicting susceptibility to social bots on twitter," in Information Reuse and Integration (IRI), 2013 IEEE 14th International Conference on, IEEE, 2013, pp. 6-13.

[71] M. Abulaish and S. Y. Bhat, "Classifier Ensembles Using Structural Features For Spammer Detection In Online Social Networks," Foundations of Computing and Decision Sciences, vol. 40, no. 2, pp. 89-105, 2015.

[72] M. S. Rahman, T.-K. Huang, H. V. Madhyastha and M. Faloutsos, "FRAppE: detecting malicious facebook applications," in Proceedings of the 8th international conference on Emerging networking experiments and technologies, ACM, 2012, pp. 313-324.

[73] M. Fire, D. Kagan, A. Elyashar and Y. Elovici, "Social Privacy Protector- Protecting Users' Privacy in Social Networks," in SOTICS 2012: Second International Conference on Social Eco--Informatics, 2012, pp. 46-50.

[74] D. Kagan, M. Fire, A. Elyashar and Y. Elovici, "Facebook Applications' Installation and Removal: A Temporal Analysis," in The Third International Conference on Social Eco-Informatics (SOTICS), Lisbon, Portugal, 2013.

[75] M. McPherson, L. Smith-Lovin and J. M. Cook, "Birds of a feather: Homophily in social networks," Annual review of sociology, pp. 415-444, 2001.

[76] N. Mishra, R. Schreiber, I. Stanton and R. E. Tarjan, "Clustering social networks," in Algorithms and Models for the Web-Graph, Springer, 2007, pp. 56-67.

[77] Y. Elovici, M. Fire, A. Herzberg and H. Shulman, "Ethical Considerations when Employing Fake Identities in OSN for Research," Science and engineering ethics, pp. 1-17, 2013.

[78] M. Fire, D. Kagan, A. Elyashar and Y. Elovici, "Friend or foe? Fake profile identification in online social networks," Social Network Analysis and Mining, vol. 4, no. 1, pp. 1-23, 2014.

[79] D. Liben-Nowell, J. Novak, R. Kumar, P. Raghavan and A. Tomkins, "Geographic routing in social networks," Proceedings of the National Academy of Sciences of the United States of America, vol. 102, no. 33, pp. 11623-11628, 2005.

## תקציר

בשנים האחרונות הרשתות חברתיות פופולריות מאוד. הן הפכו לחלק אינטגרלי וחיוני ביומיום שלנו. משתמשים מכל קצוות תבל משקיעים חלק נכבד מזמנם בגלישה לרשתות חברתיות המאפשרות להם ליצור קשרים חדשים עם משתמשים אחרים על בסיס אינטרסים משותפים, פעילויות משותפות, ורעיונות כמו גם חיזוק קשרים מן העבר.

לצד האספקטים החיוביים של הרשתות החברתיות משתמשים נאלצים להתמודד מול בעיות ייחודיות. אחת המרכזיות בהן היא פרטיות ואבטחה. פריצות אבטחה לעיתים קרובות מתרחשות כאשר משתמשים כותבים, משתפים ומפרסמים מידע פרטי לגבי עצמם, חבריהם ומקום עבודתם ברשתות החברתיות. לצערנו, לעיתים המידע אותו הם מפרסמים לציבור עלול להיות מנוצל ע"י גורמים וזדוניישישים להם אינטרס להרע. מידע פרטי נפוץ יכול להכיל תמונות, תאריך לידה, השתייכות דתית, ענייניים אישיים, דעות פוליטיות ועוד.

במחקר זה אנו מנסים להדגיש את הבעיות האקוטיות הטבועות ברשתות חברתיות החושפות את העובדים כמו גם את הארגונים להתקפות סייבר. במקרים רבים התקפות אלו כוללות שימוש בבוטים חברתיים שיכולים להפיץ ספאם ונוזקות, שימוש בפישינג וכ"ו. ההתקפות הזדוניות הללו עלולות להסתיים בגניבת זהויות, הונאות, והפסדים של נכסים אינטלקטואליים ומידע עסקי סודי.

בעיות הפרטיות והאבטחה החמורות הקשורות ברשתות חברתיות הן אלו שהיוו את ה"דלק" לשני המחקרים המשלימים הכלולים בתזה זו. במחקר הראשון אנו מפתחים אלגוריתם כללי לכריית נתונים מארגוני מטרה תוך שימוש בפייסבוק (כרגע הרשת החברתית הפופולרית ביותר בעולם) ובוטים חברתיים. באמצעות יצירת חברויות עם עובדים מארגוני מטרה הבוטים שלנו הצליחו לשחזר את המבנה הארגוני של ארגוני המטרה כמו גם לזהות קשרים חבויים ועובדים נוספים שלא היינו מסוגלים למצוא באמצעות קרולינג. אנו בחנו את השיטה המוצעת ברשת החברתית של פייסבוק והיינו מסוגלים לבנות את הרשת החברתית של העובדים בשלושה ארגונים אמיתיים שונים. בנוסף לכך, בתהליך הקרולינג עם הבוטים החברתיים האקטיביים שלנו הצלחנו לגלות עד 13.55% יותר עובדים ועד 22.27% יותר קשרים נסתרים בין עובדי הארגון לעומת תהליך הקרולינג שבוצע על ידי הבוטים הפסיביים קשרים חברים מכל סוג שהוא.

במחקר השני אנו פיתחנו אלגוריתם כללי להשגת משתמשים ספציפיים ברשתות חברתיות שהגדירו את עצמם כעובדים בארגוני המטרה תוך שימוש בטופולוגיות של הרשתות החברתיות בארגון ובבוטים חברתיים. אנו בחנו את השיטה המוצעת על משתמשי מטרה מסולמיים ארגונים אמיתים שונים בפייסבוק ושני ארגונים אמיתיים ברשת החברתית זינג (רשת חברתית גרמנית פופולרית נוספת). בסופו של דבר, הבוטים שלנו היו מסוגלים להתחבר עם משתמשים ספציפיים עם אחוזי הצלחה של עד 70% בפייסבוק ועד 60% בזינג.

התוצאות שהושגו בשני המחקרים מדגישות את הסכנות הטמונות ברשתות חברתיות. אנו מאמינים שהעלאת המודעות בייחס לנושאי פרטיות בקרב כל הישויות ברשתות החברתיות : משתמשים, ארגונים ומפעילי הרשתות כמו גם פיתוח כלים מניעתיים ופיתוח מדיניות אבטחה נוקשות יכולות לסייע במציאת פתרון בר קיימא עבור המצב הקריטי הקיים במרחב הרשתות החברתיות תוך מתן הגנה טובה יותר למשתמשי הרשתות החברתית מבחינה פרטיות ואבטחה.

# אוניברסיטת בן-גוריון בנגב, באר שבע
# הפקולטה למדעי ההנדסה
# המחלקה להנדסת מערכות מידע

## אבטחת ארגונים ויחידים ברשתות חברתיות

חיבור זה מהווה חלק מהדרישות לקבלת תואר מגיסטר בהנדסה

מאת: אביעד אלישר (aviad.elishar@gmail.com)

מנחים: פרופ׳ יובל אלוביצי׳ (elovici@post.bgu.ac.il)

ד״ר מיכאל פייר (fire@cs.washington.edu)

# אוניברסיטת בן-גוריון בנגב
# הפקולטה למדעי ההנדסה
# המחלקה להנדסת מערכות מידע

# אבטחת ארגונים ויחידים ברשתות חברתיות

חיבור זה מהווה חלק מהדרישות לקבלת תואר מגיסטר בהנדסה

מאת: אביעד אלישר
המחלקה להנדסת מערכות מידע
אוניברסיטת בן גוריון בנגב, באר שבע
טלפון: 052-6056439
דואר אלקטרוני: aviad.elishar@gmail.com

**21/09/2015**                                                                              **ח' בתשרי, התשע"ו**